\documentclass[11pt]{article}
\usepackage[margin=1in]{geometry}
\usepackage{amsmath}
\usepackage{appendix}
\usepackage{array}
\usepackage{tabularx}  
\usepackage{amsmath,amsthm,amssymb,amsfonts}
\usepackage[american]{babel}
\usepackage[url=false,giveninits=true,maxbibnames=99,style=alphabetic,maxalphanames=5]{biblatex}
\usepackage{mathtools}
\usepackage{enumitem}
\usepackage{csquotes}
\usepackage[noend]{algpseudocode}
\usepackage{algorithm}
\usepackage{xparse}
\usepackage{xspace}
\usepackage{color}
\usepackage{caption}
\usepackage{subcaption}
\usepackage{fullpage}
\usepackage{placeins}
\usepackage{xcolor}
\usepackage{makecell}
\usepackage{qcircuit}
\usepackage{thmtools}
\usepackage{thm-restate}
\usepackage[textsize=scriptsize]{todonotes}
\usepackage{verbatim}
\usepackage{qcircuit}
\usepackage{comment}
\usepackage{mathdots}
\usepackage{hyperref}
\usepackage{cleveref}

\newcommand{\red}[1]{\textcolor{red}{{#1}}}

\setuptodonotes{color=blue!15}

\mathchardef\mhyphen="2D % Define a "math hyphen"

\newcommand\newmathabbrev[2]{\newcommand{#1}{\ensuremath{#2}\xspace}}

\newcommand\cfont\mathsf
\newmathabbrev\p{\cfont{P}}
\newmathabbrev{\N}{\mathbb N}
\newmathabbrev\NP{\cfont{NP}}
\newmathabbrev\QPH{\cfont{QPH}}
\newmathabbrev\QPHpure{\cfont{pureQPH}}
\newmathabbrev\QCPH{\cfont{QCPH}}
\newmathabbrev\polyQCPH{\cfont{polyQCPH}}
\newmathabbrev\polyPH{\cfont{polyPH}}
\newmathabbrev\QEPH{\cfont{QEPH}}
\newmathabbrev\QMAH{\cfont{QMAH}}
\newmathabbrev\QAC{\cfont{QAC}_0}
\newmathabbrev\AC{\cfont{AC}}
\newmathabbrev\disagr{\mathsf{disagr}}

\newmathabbrev\SIP{\cfont{SIPSER}}

\newmathabbrev\DTIME{\cfont{DTIME}}
\newmathabbrev\tSAT{3\cfont{\mhyphen{}SAT}}
\newmathabbrev\MA{\cfont{MA}}
\newmathabbrev\AM{\cfont{AM}}
\newmathabbrev\NPDAG{\cfont{NP\mhyphen{}DAG}}
\newmathabbrev\QMADAG{\cfont{QMA\mhyphen{}DAG}}
\newmathabbrev\yes{\mathrm{yes}}
\newmathabbrev\no{\mathrm{no}}
\newmathabbrev\US{\cfont{US}}
\newmathabbrev\FP{\cfont{FP}}
\newmathabbrev\PP{\cfont{PP}}
\newmathabbrev\CeP{\cfont{C_=P}}
\newmathabbrev\coCeP{\cfont{coC_=P}}
\newmathabbrev\PH{\cfont{PH}}
\newmathabbrev\SAT{\cfont{SAT}}
\newmathabbrev\SPP{\cfont{SPP}}
\newmathabbrev\GapP{\cfont{GapP}}
\newmathabbrev\BQP{\cfont{BQP}}
\newmathabbrev\ZQEXP{\cfont{ZQEXP}}
\newmathabbrev\QP{\cfont{QP}}
\newmathabbrev\StoqMA{\cfont{StoqMA}}
\newmathabbrev\coNP{\cfont{coNP}}
\newmathabbrev\AzPP{\cfont{A_0PP}}
\newmathabbrev\QMA{\cfont{QMA}}
\newmathabbrev\QXC{\cfont{QXC}}
\newmathabbrev\QMAone{\cfont{QMA}_1}
\newmathabbrev\uQMA{\cfont{uniqueQMA}}
\newmathabbrev\uSAT{\cfont{uniqueSAT}}
\newmathabbrev\cQMA{\cfont{cloneableQMA}}
\newmathabbrev\coQMA{\cfont{coQMA}}
\newmathabbrev\BPP{\cfont{BPP}}
\newmathabbrev\QCMA{\cfont{QCMA}}
\newmathabbrev\pNPlog{\p^{\NP[\log]}}
\newmathabbrev\pNP{\p^{\NP}}
\newmathabbrev\pNPtwo{\p^{\NP[2]}}
\newmathabbrev\pNPone{\p^{\NP[1]}}
\newmathabbrev\pParSAT{\p^{||\SAT}}
\newmathabbrev\pQMApar{\p^{||\QMA}}
\newmathabbrev\pCpar{\p^{||\C}}
\newmathabbrev\pStoqMApar{\p^{||\StoqMA}}
\newmathabbrev\pQMAlog{\p^{\QMA[\log]}}
\newmathabbrev\pClog{\p^{\textup{C}[\log]}}
\newmathabbrev\pC{\p^{\textup{C}}}
\newmathabbrev\QMASPACE{\cfont{QMASPACE}}
\newmathabbrev\pQMAtlog{\p^{\QMA(2)[\log]}}
\newmathabbrev\pStoqMAlog{\p^{\StoqMA[\log]}}
\newmathabbrev\pQMApt{\p^{\Vert\QMA(2)}}
\newmathabbrev\pQMA{\p^{\QMA}}
\newmathabbrev\SharpP{\cfont{\#P}}
\newmathabbrev\pSharP{\p^{\SharpP[1]}}
\newmathabbrev\PromisePP{\cfont{PromisePP}}
\newmathabbrev\lett{\le_\mathrm{tt}}
\newmathabbrev\YES{\mathsf{YES}}
\newmathabbrev\NO{\mathsf{NO}}
\newmathabbrev\PSPACE{\cfont{PSPACE}}
\newmathabbrev\IP{\cfont{IP}}
\newmathabbrev\POLY{\cfont{POLY}}
\newmathabbrev\DAG{\cfont{DAG}}
\newmathabbrev\StoqMADAG{\StoqMA\mhyphen\cfont{DAG}}
\newmathabbrev\CDAG{C\mhyphen\cfont{DAG}}
\newmathabbrev\CDAGf{C\mhyphen\cfont{DAG}_f}
\newmathabbrev\CDAGs{C\mhyphen\cfont{DAG}_s}
\newmathabbrev\CDAGd{C\mhyphen\cfont{DAG}_{d}}
\newmathabbrev\CDAGo{C\mhyphen\cfont{DAG}_1}
\newmathabbrev\LOGS{\cfont{LOGS}}
\newmathabbrev\TAUT{\cfont{TAUTOLOGY}}
\newmathabbrev\SBQP{\cfont{SBQP}}
\newmathabbrev\Fc{F_\coNP}
\newmathabbrev\Fa{F_\AzPP}
\newmathabbrev\GSCON{\cfont{GSCON}}
\newmathabbrev\GSCONexp{\GSCON_\cfont{exp}}
\newmathabbrev\QMAexp{\QMA_\cfont{exp}}
\newmathabbrev\UQMA{\cfont{UQMA}}
\newmathabbrev\R{\mathbb R}
\newmathabbrev\Trees{\cfont{TREES}}
\newmathabbrev\apxsim{\cfont{APX\mhyphen{}SIM}}
\newmathabbrev\AWPP{\cfont{AWPP}}
\newmathabbrev\X{\mathcal{X}}
\newmathabbrev\Y{\mathcal{Y}}

\newmathabbrev\Z{\mathcal{Z}}

\newmathabbrev\ZZ{\mathbb{Z}}
\newmathabbrev\Hprop{H_\mathrm{prop}}
\newmathabbrev\Hin{H_\mathrm{in}}
\newmathabbrev\Hout{H_\mathrm{out}}
\newmathabbrev\Hstab{H_\mathrm{stab}}
\newmathabbrev\Lext{\L_\mathrm{ext}}
\newmathabbrev\BTWNP{\cfont{BTW}(\NP)}
\newmathabbrev\BSN{\cfont{BSN}}
\newmathabbrev\SN{\cfont{SN}}
\newmathabbrev\BD{\cfont{BD}}
\newmathabbrev\HYPERTREE{\cfont{NP\mhyphen{}HYPERTREE}}
\newmathabbrev\Hext{H_\mathrm{ext}}
\newmathabbrev\Hpropt{\tilde{H}_\mathrm{prop}}
\newmathabbrev\Hint{\tilde{H}_\mathrm{in}}
\newmathabbrev\Houtt{\tilde H_\mathrm{out}}
\newmathabbrev\EXP{\cfont{EXP}}
\newmathabbrev\A{\mathcal{A}}
\newmathabbrev\U{\mathcal{U}}

\renewcommand\L{\mathcal{L}}

\newmathabbrev\DAGSSAT{\DAGS(\SAT)}
\newmathabbrev\DAGS{\mathrm{DAGS}}
\newmathabbrev\DAGSNP{\DAGS(\NP)}

\newmathabbrev\AND{\cfont{AND}}

\newmathabbrev\STCONN{{S,T}\cfont{\mhyphen{}CONN}}
\newmathabbrev\CNF{\cfont{CNF}}
\newmathabbrev\NEXP{\cfont{NEXP}}
\newmathabbrev\NPSPACE{\cfont{NPSPACE}}
\newmathabbrev\QCMASPACE{\cfont{QCMASPACE}}
\newmathabbrev\BQPSPACE{\cfont{BQPSPACE}}
\newmathabbrev\PQPSPACE{\cfont{PQPSPACE}}
\newmathabbrev\PQP{\cfont{PQP}}
\newmathabbrev\TQBF{\cfont{TQBF}}
\newmathabbrev{\PCP}{\cfont{PCP}}
\newmathabbrev\BQUPSPACE{\cfont{BQ_UPSPACE}}
\newmathabbrev\QMAt{\QMA(2)}
\newmathabbrev\QMAtSEP{\QMA^{\mathsf{SEP}}(2)}
\newmathabbrev\QMAtexp{\QMAt_{\exp}}
\newmathabbrev\MIP{\cfont{MIP}}
\newmathabbrev\MIPt{\MIP(2)}
\newmathabbrev\BellQMA{\cfont{BellQMA}}
\newmathabbrev\BellQMAt{\BellQMA(2)}
\newmathabbrev\BellQMAtexp{\BellQMAt_{\exp}}

\protected\def\verythinspace{%
  \ifmmode
    \mskip0.5\thinmuskip
  \else
    \ifhmode
      \kern0.08334em
    \fi
  \fi
}

\newcommand{\C}{\mathbb C}

\newcommand{\be}{\begin{equation}}
\newcommand{\ee}{\end{equation}}

\renewcommand{\epsilon}{\varepsilon}

\DeclarePairedDelimiterX\braket[2]{\langle}{\rangle}{#1 \delimsize\vert #2}
\DeclarePairedDelimiterX\ketbra[2]{\lvert}{\rvert}{#1 \delimsize\rangle\delimsize\langle #2}

\setlist[itemize]{noitemsep, topsep=0pt}
\setlist[enumerate]{noitemsep, topsep=0pt}

\declaretheorem[numberwithin=section]{theorem}

\declaretheorem[sibling=theorem,style=definition]{definition}

\crefname{observation}{observation}{observations}
\Crefname{observation}{Observation}{Observations}

\makeatletter
\newcommand{\subalign}[1]{%
  \vcenter{%
    \Let@ \restore@math@cr \default@tag
    \baselineskip\fontdimen10 \scriptfont\tw@
    \advance\baselineskip\fontdimen12 \scriptfont\tw@
    \lineskip\thr@@\fontdimen8 \scriptfont\thr@@
    \lineskiplimit\lineskip
    \ialign{\hfil$\m@th\scriptstyle##$&$\m@th\scriptstyle{}##$\hfil\crcr
      #1\crcr
    }%
  }%
}
\NewDocumentCommand{\LeftComment}{s m}{%
  \Statex \IfBooleanF{#1}{\hspace*{\ALG@thistlm}}\(\triangleright\) #2}

\def\moverlay{\mathpalette\mov@rlay}
\def\mov@rlay#1#2{\leavevmode\vtop{%
   \baselineskip\z@skip \lineskiplimit-\maxdimen
   \ialign{\hfil$\m@th#1##$\hfil\cr#2\crcr}}}
\newcommand{\charfusion}[3][\mathord]{
    #1{\ifx#1\mathop\vphantom{#2}\fi
        \mathpalette\mov@rlay{#2\cr#3}
      }
    \ifx#1\mathop\expandafter\displaylimits\fi}

\makeatother

\algnewcommand{\LineComment}[1]{\State \(\triangleright\) #1}

\algblockdefx[ON]{Blk}{EndBlk}[1]
  {#1}
  {}

\makeatletter
\ifthenelse{\equal{\ALG@noend}{t}}%
  {\algtext*{EndBlk}}
  {}%
\makeatother

\AtEveryBibitem{%
  \clearlist{language}%
}

\def\({\left(}
\def\){\right)}
\def\X{\mathcal{X}}
\def\Y{\mathcal{Y}}
\def\Z{\mathcal{Z}}

\def\yes{\text{yes}}
\def\no{\text{no}}

\usepackage{authblk}
\usepackage{longtable} 
\usepackage{booktabs}  
\usepackage{seqsplit}
\renewcommand{\arraystretch}{0.9}
\makeatletter
\renewcommand\@makefnmark{\hbox{\textsuperscript{\@thefnmark}}}
\makeatother

\usepackage{amssymb}
\usepackage{tikz}
\usetikzlibrary{calc}
\usetikzlibrary{positioning}
\usepackage{hyperref}
\usepackage{xurl}

\hypersetup{
colorlinks = true,
citecolor= blue,
linkcolor= blue,
breaklinks=true
}
\title{\textsf{On Error Thresholds 
for Pauli Channels} \\ 
\textsf{\Large Some answers with many more questions}}

\date{}

\author[1,2]{Avantika Agarwal}
\author[1,3]{Alan Bu}
\author[1,2,4]{Amolak Ratan Kalra}
\author[1,4,5]{Debbie Leung}
\author[1,2]{Luke Schaeffer}
\author[1,6]{Graeme Smith}

\affil[1]{Institute for Quantum Computing, University of Waterloo}
\affil[2]{David R. Cheriton School of Computer Science, University of Waterloo}
\affil[3]{Department of Physics, Harvard University}
\affil[4]{Perimeter Institute for Theoretical Physics}
\affil[5]{Department of Combinatorics and Optimization, University of Waterloo}
\affil[6]{Department of Applied Mathematics, University of Waterloo}

\begin{document}

\maketitle

\begin{abstract}

This paper focuses on error thresholds for Pauli channels. We numerically compute lower bounds for the thresholds using the analytic framework of coset weight enumerators pioneered by DiVincenzo, Shor and Smolin in 1998. In particular, we study potential non-additivity of a variety of small stabilizer codes and their concatenations, and report several new concatenated stabilizer codes of small length that show significant non-additivity. We also give a closed form expression of coset weight enumerators of concatenated phase and bit flip repetition codes. Using insights from this formalism, we estimate the threshold for concatenated repetition codes of large lengths. Finally, for several concatenations of small stabilizer codes we optimize for channels which lead to maximal non-additivity at the hashing point of the corresponding channel. We supplement these results with a discussion on the performance of various stabilizer codes from the perspective of the non-additivity and threshold problem. We report both positive and negative results, and highlight some counterintuitive observations, to support subsequent work on lower bounds for error thresholds.
\end{abstract}

\clearpage

\section{Introduction}

In 1948, Shannon in his landmark paper \cite{shannon} derived a formula for the capacity of an arbitrary noisy communication channel in terms of the mutual information of the input and output, maximized over the input distribution.
To do this, he showed that this optimized expression is an upper bound on the capacity, and that this rate is achievable using a random code.  
Furthermore, this optimized expression, when evaluated on two channels used in parallel, has a joint maximum that is simply the sum of the individual maxima; correspondingly, the optimization problem is said to exhibit a property called \emph{additivity}.  
Shannon's work establishes that the capacity of the binary symmetric channel is $1-h$ where $h$ is the binary entropy of the probability of error.  

The quantum analogue of the capacity problem turns out vastly more intricate.  
The complexity is fully manifest already on the qubit depolarizing channel which generalizes the binary symmetric channel. In \cite{bennett1996mixed}, a random stabilizer code\footnote{Reference \cite{bennett1996mixed} establishes the equivalence between entanglement purification and quantum error correction.  We adopt the latter framework throughout this paper.  The random hashing method for entanglement purification in \cite{bennett1996mixed} is translated to a randomized quantum error correcting code subsequently commonly called a random stabilizer code \cite{Gottesman96}.} \cite{Gottesman96}
is found to achieve the rate $1-h'$ where $h'$ is the entropy of the Pauli errors.  Furthermore, this rate also matches the upper bound given by the quantum Hamming bound \cite{Gottesman96} for correcting the typical errors using non-degenerate codes. 
One might thus expect this result to provide the proper quantum generalization of the capacity of the depolarizing channel.  So it was a big surprise when the authors in \cite{shor1996quantumerrorcorrectingcodesneed} showed that one could outperform random codes by first applying a small degenerate code to the noisy channel, before concatenating with a random stabilizer code.  Furthermore, they showed that this method achieves a positive rate when $h'$ is slightly above $1$.  
Consequently, the quantum capacity cannot always be attained by random codes.  
 
The novel coding technique in \cite{shor1996quantumerrorcorrectingcodesneed} was further developed in \cite{divincenzo1998quantum} while a series of work \cite{Schumacher96,SchumacherNielsen96,lloyd1997capacity,cerf1998quantum,PhysRevA.57.4153,850671} provided crucial understanding of the quantum capacity of an arbitrary noisy quantum channel, culminated in a quantum channel capacity theorem 
\cite{lloyd1997capacity,shor2002quantum,devetak2005private} 
with an expression for the capacity of a quantum channel based on the \emph{coherent information} of the output of the channel maximized over all input states.  
However, unlike the classical setting, the maximized coherent information is not additive, and we need to maximize the coherent information jointly over many uses ($n$) of the channel, and divide by $n$ to obtain the capacity expression.  This process is called regularization, and it prevents a closed form expression for the capacity to be obtained for most quantum channels.  
Furthermore, there is no known general algorithm to determine if the quantum capacity of a given channel is zero or positive, and we do not even have an upper bound on the value of $n$ for the regularization \cite{cubitt2015unbounded}.  

Using the capacity formula, the previous results \cite{shor1996quantumerrorcorrectingcodesneed, divincenzo1998quantum} can be interpreted as code examples that exhibit non-additivity of coherent information. Furthermore, a variety of such non-additivity results have been shown since \cite{smith2007degenerate,smith2008quantum,gsquantum,fern, smith2011quantum, brandao2012does, cubitt2015unbounded, jackson2017degenerate, leditzky2018dephrasure, bausch2020quantum, bausch2021error, leditzky2023generic, bhalerao2025improvingquantumcommunicationrates}.  
However, they are more a menagerie of quantum non-additivity effects, providing hints but not a theory on what quantum error correcting codes are capacity achieving.  
Even more surprisingly, in spite of these results, we still do not know the maximum error rates at which many natural quantum channels (such as the depolarizing channel) have positive capacity.  This problem is sometimes referred to as the ``threshold problem'' in channel capacity.  Given the capacity formula and its continuity on the error rates, finding better lower bounds on threshold boils down to finding codes which show large non-additivity. In this paper, we study non-additivity and the threshold problem for Pauli channels (see next section for definition) using the framework of coset weight enumerators \cite{divincenzo1998quantum}.

\subsection{Previous Work}

This problem was previously studied in multiple works. In this section, we review these results and summarize what is known about the problem. 

Throughout, the probability vector $(p_I, p_X, p_Y, p_Z)$ is associated with the mixed Pauli channel ${\cal N}(\rho) = p_I \rho + p_X X \rho X + p_Y Y \rho Y + p_Z Z \rho Z$ where $X,Y,Z$ are the qubit Pauli operators.  The entropy of this probability vector is also called the entropy of the channel.  

In 1996, Bennett, DiVincenzo, Smolin and Wootters \cite{bennett1996mixed} showed that 
using a random stabilizer code (equivalent to random hashing of EPR pairs), one could achieve the rate $1$ minus the channel entropy, thus the capacity is positive until the channel entropy is $1$. For the $(1-3p, p, p, p)$ depolarizing channel, this gives non-zero capacity up to $p=0.063096$.  Concurrently, Shor and Smolin \cite{shor1996quantumerrorcorrectingcodesneed} showed that encoding with the 5-qubit phase-flip repetition code (and concatenated with random stabilizer code) leads to positive rate up to $p=0.063466$.  In 1998, this was further improved by DiVincenzo, Shor and Smolin \cite{divincenzo1998quantum}, where a $5\times 5$ concatenated bit- and phase-flip repetition code gives positive rate up to $p=0.06352$. 
(A $k \times l$ concatenated bit- and phase-flip repetition code first encodes the logical qubit into an $l$-qubit phase-flip repetition code, and then encoding each of the $l$ qubits into a $k$-qubit bit-flip repetition code, before transmitting each of the $lk$ qubits via the noisy channel.)  The code in \cite{divincenzo1998quantum} first encodes the logical qubits with the random stabilizer code, and then encodes each qubit in the random code by the $k \times l$ repetition code.  To avoid cumbersome writing, we omit mentioning the concatenation with random stabilizer code most of the time, but this step is always needed to achieve the rate.   
In 2007, relying on the idea of degeneracy of repetition codes, Smith and Smolin \cite{smith2007degenerate} furthered the analysis and their proposed $5\times 16$ concatenated phase- and bit-flip repetition code gave a positive rate up to the improved $p=0.063626$. The authors also initiated the study of the threshold problem for other related Pauli channels such as the independent X-Z channel with error probabilities $((1-p)^2, p(1-p), p^2, p(1-p))$ and the 2-Pauli channel with error probabilities $(1-2p, p, 0, p)$.

In 2008, Fern and Whaley \cite{fern} improved the lower bound on the error threshold for all three aforementioned Pauli channels.  Their Monte Carlo numerical method applies to many levels of concatenation of small codes of very large blocklengths.  
In particular, 
they find that adding several self-concatenations of the distance-$3$ $5$-qubit code between the random stabilizer code and the concatenated bit- and phase-flip repetition code gives positive rates up to $p=0.063766$ for the depolarizing channel, $p=0.11398$ for 2-Pauli channel and $p=0.11277$ for independent X-Z channel. One should note that these codes encode a single qubit into $\sim 5^{10}$ qubits. If one restricts attention to concatenated repetition codes, their best codes include the $5\times 51$ repetition code for the depolarizing channel (positive rate up to $p=0.0637338273$), the $5\times 74$ repetition code for the 2-Pauli channel (positive rate up to $p=0.1139425214$), and the $5\times 77$ repetition code for the independent X-Z channel (positive rate up to $p=0.1127458434$). 

Very recently, in \cite{bhalerao2025improvingquantumcommunicationrates} 
permutation invariant codes were used to study the threshold problems.  
These codes improve the lower bounds for the thresholds for 2-Pauli and independent X-Z channels, and the optimal blocklengths are $n=24$ and $n=18$ respectively. 
These are not stabilizer codes, and the symmetry of the codes enabled the analysis of the medium blocklengths.  These codes do not improve the lower bound for the threshold for the depolarizing channel.  In spite of all the results mentioned, and some more recent work \cite{fan2024overcoming,bausch2021error,bausch2020quantum}, the question of what quantum codes can best extend the lower bound for the threshold is not well understood.

\subsection{Our Results}
We study the thresholds and non-additivity for long concatenated repetition codes and a number of small stabilizer codes and their concatenations. We also optimize for Pauli channels that lead to maximal coherent information. In particular, we have the following results:
\begin{itemize}
\item We perform a search over several small stabilizer (See Tables \ref{table1}, \ref{table7}) and concatenated stabilizer codes (See Tables \ref{table2}, \ref{table3}, \ref{table4}, \ref{table5}, \ref{table8}) and find several new codes that perform better than random stabilizer codes both for depolarizing and independent X-Z channel. The codes that we study are concatenations of repetition codes, $[[4,2,2]]$ code, 5-qubit code, 7-qubit code, the $[[8,2,2]]$ toric code, the 9-qubit code for biased error proposed in \cite{biased} and the holographic codes proposed in \cite{fan2024overcoming}. The list of all stabilizer codes that we used is given in Appendix \ref{app} together with their generators and logical operators.

\item We find that concatenating the 5-repetition code with the biased 9-qubit code leads to better thresholds\footnote{For a fixed channel, each code provides a positive rate when concatenated with a random stabilizer code up to some noise parameter.  The noise level when the rate just turns zero will be called the threshold for the code.  This is a lower bound for the threshold noise level for the channel.} 
than just a single layer of 5 repetition code for both depolarizing and independent X-Z channel (see Tables \ref{table2}, \ref{table8}).
\item We find that concatenating repetition codes with holographic codes generally leads to improved thresholds for both depolarizing and independent X-Z channel (See Tables \ref{table3}, \ref{table8}). Note that the 5-qubit code is also an example of a holographic code \cite{pastawski2015holographic} and concatenating repetition codes with multiple layers of it achieved the best threshold for depolarizing channel \cite{fern}.
\item We estimate the threshold for concatenations of long repetition codes of lengths upto 15 $\times$ 7000 for all three Pauli channels (See Figures \ref{depolarizingplot}, \ref{fig:indxzplot} and \ref{fig:twopauliplot}). This is a significantly larger search space than previously explored. We find that increasing the length of the first repetition code generally leads to worse thresholds. Increasing the length of the second repetition code shows an improvement in threshold up to a certain length and then gets worse, this pattern was previously seen for small length repetition codes. However, long concatenated repetition codes are the only examples we could find which do better than hashing for the 2-Pauli channel.
\item The main result that enables such a search is to use the coset weight enumerator framework, first proposed in this context in \cite{divincenzo1998quantum}. We also give a closed form expression for the coset weight enumerators of concatenated repetition codes in Section \ref{weightenumerators}. 

\item To further understand the structure of concatenations that improve the threshold, we study different combinations of concatenated codes. In particular, we find that 3 layers of repetition codes do worse than a single layer of repetition code (as reported in Table \ref{table4}). So continuing to concatenate repetition codes should not lead to improved thresholds. In the particular case of concatenating the repetition codes with the 5-qubit code, we find that putting the 5-qubit code in the middle layer gives better threshold than putting it at the end. Finally, we find that for $n\leq 7$, when concatenating repetition codes with permutation codes, the best permutation codes are repetition codes themselves for the depolarizing channel.

\item For a number of concatenated stabilizer codes, we fix the code and vary the underlying Pauli channel and find the channel that leads to maximal coherent information at the corresponding hashing point (when the rate is zero for random stabilizer code).  
Note that the coherent information is also the quantitative amount of non-additivity demonstrated.  
(See Tables \ref{table10} and \ref{table11}). We find that even after the optimization, some codes do not show non-additivity. Most codes show highest coherent information for highly biased Pauli channels and a few of them show highest coherent information for channels with roughly equal probabilities of two errors. These results are also of independent interest from the perspective of quantum error correction in the presence of biased noise.

\item If we have two codes C1 and C2 such that C1 has better threshold than C2, the ordering of performance need not be preserved when each of C1 and C2 is concatenated with a third code C3. In particular, it is possible that
\begin{align*}
    \mathsf{threshold}(C_1) > \mathsf{threshold}(C_2) \text{ but } \exists~C_3 \text{ such that } \mathsf{threshold}(C_1 \times C_3) < \mathsf{threshold}(C_2 \times C_3)
\end{align*}

For example, when C1 is 7-repetition, C2 is 3-repetition, and C3 is 7-repetition, we indeed see such reversal of performance.

This mysterious phenomenon vastly complicates our study, because this means we cannot optimize for codes layer by layer.
\end{itemize}

\textbf{Note:} All of the code, interactive plots for Figures \ref{depolarizingplot}, \ref{fig:indxzplot} and \ref{fig:twopauliplot} and a spreadsheet with all the data is available on GitHub \cite{github}. The list of all stabilizer codes that we used is given in Appendix \ref{app} together with their generators and logical operators. The tables presented throughout the paper are sorted in descending order of thresholds.

\subsection{Some Lessons}
In this subsection, we highlight some counterintuitive observations from our results that illustrates the obstacles to finding codes with improved thresholds. These are discussed in detail when we present the corresponding numerical results.
\begin{itemize}
    \item {\bf Degeneracy over distance.} 
    Every code with non-trivial distance (e.g., the 5-qubit and the 7-qubit code), used as a single layer, has a threshold worse than using random stabilizer codes alone.  
    This holds true even after concatenating them with a repetition code in the second layer.
    This observation is totally at odds with the fact that the distance is roughly twice the number of correctable errors, and challenges our intuition that noisy channel coding is a specific application of error correction.  
    Entropy reduction seems to dominate the threshold problem, not quite correcting errors, until we reach a higher level of the concatenated code.   
    \item {\bf Tailor-made codes help.} The threshold is improved by concatenating repetition codes with codes tailored to correct Pauli channels with very biased errors.  So, tailoring the codes in the second layer to target the typical effective channels of the first layer might be a good strategy. 
    \item {\bf Too many layers? More might not be merrier!} 
    Given the success of the concatenated bit- and phase-flip repetition codes of short block-lengths in the threshold problem, one might be tempted to believe that alternating many layers of the two types of repetition codes is a good strategy, but we find the opposite.  Three or more layers of repetition codes gives a threshold worse than a single layer!  Three layers also performs worse than swapping the mid-layer with the 5-qubit code.     
    \item {\bf Long repetition codes crash and burn.} As you increase the length of the concatenated repetition codes, the threshold becomes worse overall. On increasing the length of the second repetition code after fixing the first one, the threshold improves up to a certain length and then starts to get worse again.  Intuitively, the error correcting ability of one type of error is enhanced at the expense of spreading the other error more, but there is no operational understanding of these optimal block-lengths.  
    \item {\bf Diversity might be key!}  
    Another appealing idea is to take the $5 \times 51$ bit- and phase-flip repetition code and concatenate a third repetition code to catch a potentially really really long optimal block-length for the theshold. But we see good thresholds using only up to two layers of short repetition code and some other codes, and these are more efficient in practice and more tractable to analyse.  
    \item {\bf Permutation invariant versus stabilizer codes.} The permutation invariant codes in \cite{bhalerao2025improvingquantumcommunicationrates} improves the thresholds for the 2-Pauli and independent X-Z channel very similarly, but not that of the depolarizing channel.  Meanwhile, the concatenated stabilizer codes we study improves the thresholds for the depolarizing channel and independent X-Z channel very similarly, and the threshold of 2-Pauli channel responds to the codes very differently.  In summary, for 2-Pauli channel, permutation invariant codes do well but stabilizer codes don't. For independent X-Z channel, both stabilizer codes and permutation invariant codes do well. For depolarizing channel, stabilizer codes do well.

    \item {\bf Best small permutation invariant codes = repetition codes.} On concatenating repetition codes with permutation invariant codes for the depolarizing channel, we find that for small block-lengths, the best permutation invariant codes are repetition codes.
    \item {\bf Non-additivity is everywhere!}  We find that non-additivity is a rather general phenomenon, in the sense that for most stabilizer codes, there is a Pauli channel where the code exhibits non-additivity.

\end{itemize}

\section{Preliminaries}
\subsection{Setup and Construction}
\begin{figure}
    \centering
    \includegraphics[width=0.55\linewidth]{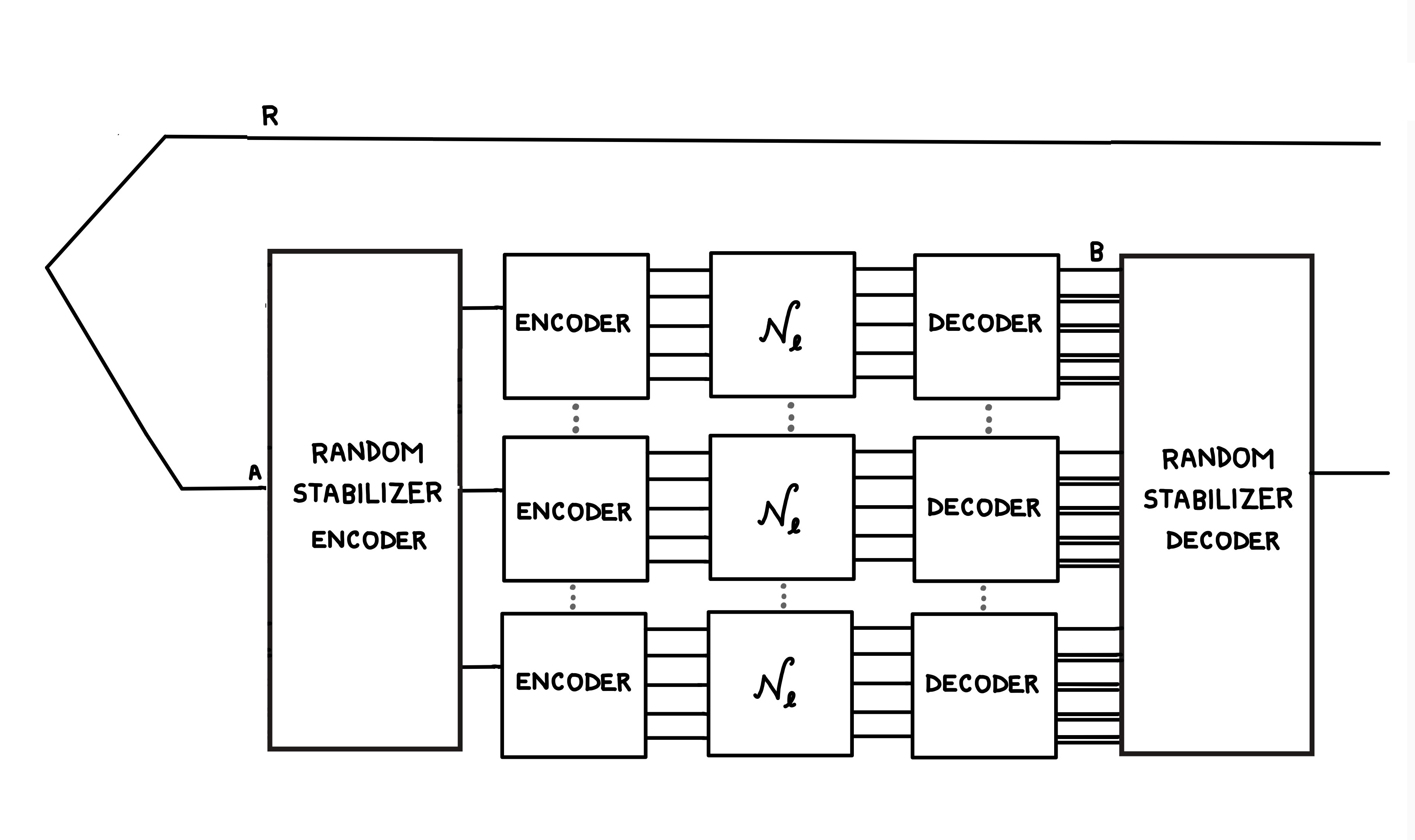}
    \caption{The depiction of the channel capacity setup \cite{divincenzo1998quantum}. The input is first encoded using a random stabilizer code, and each output qubit is further encoded by a code C. The final encoded qubits pass through the channel $\mathcal{N}_l = \mathcal{N}^{\otimes l}$. The decoder then performs syndrome measurements of C followed by those of the random stabilizer code.}
    \label{fig1}
\end{figure}

In this section, we review the setup of \cite{divincenzo1998quantum} for computing a lower bound for the error threshold of the depolarizing channel (See Fig \ref{fig1}). Bennett, DiVincenzo, Smolin, and Wootters in \cite{bennett1996mixed} gave the first lower bound on error threshold using a random stabilizer code on $n$ qubits (with $n-k$ random stabilizer generators), where $n$ is the number of uses of the depolarizing channel. In particular, they showed that with high probability over the choice of a random stabilizer code, every syndrome measurement reduces the entropy of the system by 1 bit. Since there are roughly 
$2^{n\mathsf{H}(1-3p, p, p, p)}$ typical Pauli errors for the depolarizing channel, one can identify and correct the Pauli error with 
$\approx n\mathsf{H}(1-3p, p, p, p)$ syndrome measurements.  Here, $\mathsf{H}$ is the entropy function. Therefore, the rate $1-\mathsf{H}(1-3p, p, p, p)$ is achievable, and the error threshold $p^{\ast}$ is the value of $p$ for which $\mathsf{H}(1-3p, p, p, p) = 1$, this is approximately $p^{\ast} = 0.063096$.

However, it was shown in \cite{shor1996quantumerrorcorrectingcodesneed, divincenzo1998quantum} that it is possible to achieve a positive rate at higher error rates. They do this by further encoding each of the $n$ qubits of the random stabilizer code using a $5$-qubit repetition code and then passing the resulting $5n$ encoded qubits through the depolarizing channel, to get a positive rate up to $p=0.063466$. After this, the circuit measures the syndrome for each of the $5$-qubit repetition codes to get back $n$ qubits. After this $n-k$ measurements for the stabilizer generators of the random stabilizer code are performed to identify the error, which is then reverted.  See Figure \ref{fig1} for the circuit. Finally, \cite{shor1996quantumerrorcorrectingcodesneed, divincenzo1998quantum} gave an expression for the rate in terms of the von Neumann entropy of the system $RB$.
\begin{align}\label{eq:capacity}
    Q = \frac{1}{l} (1-S_{RB})
\end{align}
where $l$ is the length of the code used for encoding the state, which is 5 in this case.

\begin{figure}
    \centering
    \includegraphics[width=0.50\linewidth]{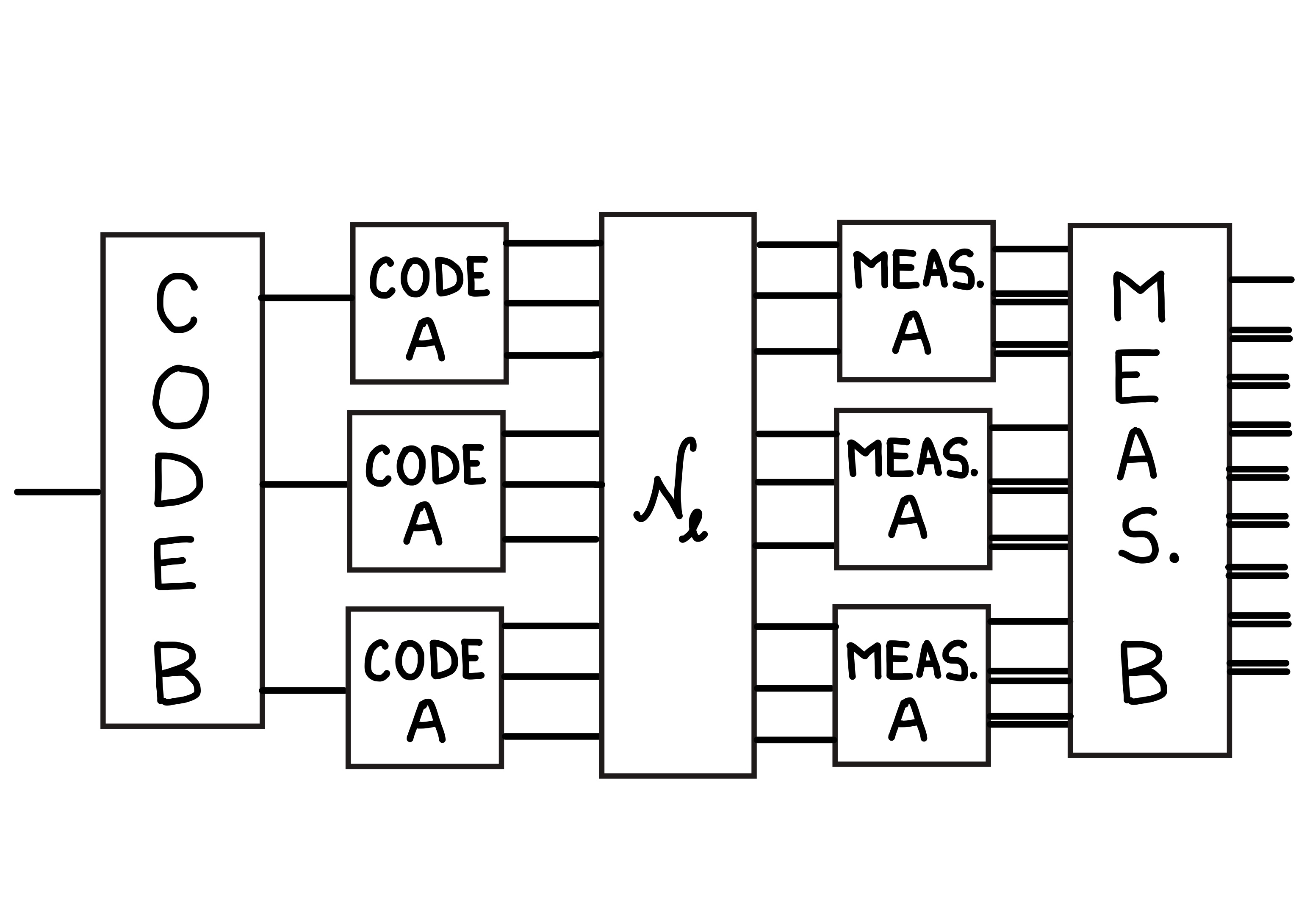}
    \caption{Order of code concatenation}
    \label{fignotation}
\end{figure}

\paragraph{Notation.} We use the following notational convention to refer to concatenated codes (see Figure \ref{fignotation} for the circuit):
\begin{enumerate}
    \item We use Code $A \times \text{Code}~B$ to refer to the following order of concatenation: first encode the state using Code $B$, then encode the output qubits using Code $A$. The resulting qubits are then passed through the channel. After the action of the channel, first the syndrome measurement for Code $A$ is performed, then for Code $B$ and finally random hashing is done on the remaining qubits.
    \item In words, we use the notation ``Code $A$ concatenated with Code $B$'' to refer to the order of concatenation stated above.
    \item We use the notation ``first'' layer to refer to Code $A$ and ``second'' layer to refer to Code $B$ in the order of concatenation stated above.
\end{enumerate}

\subsection{Review of Weight Enumerators}\label{sec:wtenumrev}
In this section, we will describe how to use the coset weight enumerator framework to compute the entropy of an encoded state after transmission through the depolarizing channel \cite{divincenzo1998quantum}, which in turn enables us to compute the error threshold. Weight enumerators polynomials have been used on many occasions before, both in quantum error correction and quantum information theory, see \cite{calderbank1998quantum, rains1, rains2, rains1999rigorous,divincenzo1998quantum, gottesman2024surviving, kalra2025invariant}. We start by recalling the definition of the weight enumerator for a stabilizer code. Note that in the following discussion, we only consider the Pauli group without phases.
\begin{definition}[Weight Enumerator \cite{gottesman2024surviving}]
A weight enumerator for an $[[n,k,d]]$ quantum stabilizer code with the set $S$ of stabilizers is a homogeneous polynomial $W(x,y)$ of degree $n$ which is defined as follows:
\begin{align}
W(x,y)=\sum_{M\in S}x^{n-\mathsf{wt}(M)}y^{\mathsf{wt}(M)}
\end{align}
Here $\mathsf{wt}(M)$ is the number of non-identity Pauli operators in $M$. In particular, the coefficient of $x^{n-k}y^{k}$ is the number of stabilizers of weight $k$.
\end{definition}
Note that if we evaluate $W(x,y)$ at $x=1-3p$, $y=p$ we get the probability that the $(1-3p, p, p, p)$ depolarizing channel applied an error $M\in S$ \cite{shor1997quantum}. We can also define similar homogeneous polynomials that capture the weight distribution of arbitrary subset $T\subseteq E$ of Pauli errors on $n$ qubits as follows:
\begin{align}
P_{T}(x,y)= \sum_{M\in T}x^{n-\mathsf{wt}(M)}y^{\mathsf{wt}(M)}
\end{align}
We can use such polynomials to compute the different probabilities used to calculate the entropy for Equation \ref{eq:capacity}. In particular, for any two Pauli error $M_{1}$ and $M_{2}$ their syndrome is the same iff $M_{1}^{\dagger}M_{2}\in N(S)$. This means that every syndrome $s$ corresponds to a distinct coset in $E/N(S)$. For a particular coset representative $M\in \overline{E}$ we denote the corresponding normalizer coset $[M]_{N(S)}$. Then the probability that we see the corresponding syndrome $s_{M}$ is equal to $P_{[M]_{N(S)}}(1-3p, p)$. 

\begin{figure}
    \centering
    \includegraphics[width=0.4\linewidth]{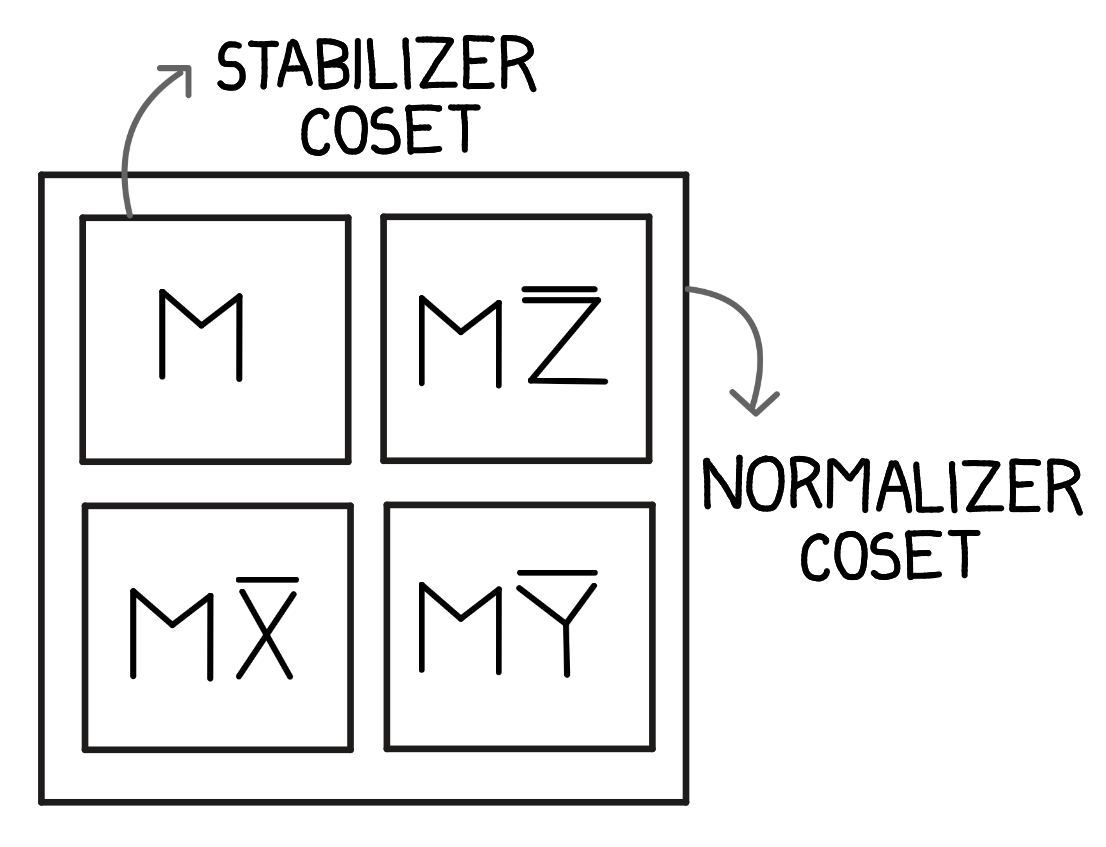}
    \caption{Normalizer coset partitioned by stabilizer cosets}
    \label{coset}
\end{figure}

Further, for any two errors $M_1, M_2$, the output state is the same iff $M_1^{\dagger}M_2 \in S$. Thus the output state can be characterized using distinct cosets in $E/S$. For a particular coset representative $M \in E$, we denote the corresponding stabilizer coset $[M]_{S}$. Note that $[M]_{N(S)} = [M]_S \sqcup [M\overline{Z}]_S \sqcup [M\overline{X}]_S \sqcup [M\overline{Y}]_S$, where the four stabilizer cosets partition the normalizer coset, and $\overline{X}, \overline{Y}, \overline{Z}$ denote logical $X, Y, Z$ operators (see Figure \ref{coset}). Moreover, for any errors $M_1, M_2$ from distinct partitions, the resulting states are orthogonal, since $\langle \Phi | I \otimes (M_1^{\dagger}M_2)|\Phi\rangle = 0$. Therefore, we can compute the von Neumann entropy of the state resulting from a syndrome $s_M$ by computing the Shannon entropy of the conditional probabilities that the depolarizing channel applies an error from the distinct partitions of $[M]_{N(S)}$. The probability that the depolarizing channel applies an error from a particular partition $[M]_S$ is $P_{[M]_S}(1-3p, p)$.

For a particular normalizer coset $T \in E/N(S)$, suppose we pick a coset representative $M_T$. Then we denote the stabilizer cosets $[M_T]_S, [M_T\overline{Z}]_S, [M_T\overline{X}]_S,[M_T\overline{Y}]_S$ by $T_{\overline{I}}, T_{\overline{Z}}, T_{\overline{X}}, T_{\overline{Y}}$ respectively. Then, we can compute the entropy $S_{RB}$ from Equation \ref{eq:capacity}, using weight enumerators by the following expression:
\begin{align}\label{eq:ent}
\sum_{T \in E/N(S)}P_{T}(1-3p, p) H\left(\frac{P_{T_{\overline{I}}}(1-3p, p)}{P_{T}(1-3p, p)}, \frac{P_{T_{\overline{X}}}(1-3p, p)}{P_{T}(1-3p, p)}, \frac{P_{T_{\overline{Y}}}(1-3p, p)}{P_{T}(1-3p, p)}, \frac{P_{T_{\overline{Z}}}(1-3p, p)}{P_{T}(1-3p, p)}\right)
\end{align}
In order to study the threshold probability for a channel with independent $X$ and $Z$ noise, which applies $X$ and $Z$ error independently with probability $p$ each, we need to distinguish between the $X,Y,Z$ components of an error since they occur with different probabilities. A useful tool to do that is the complete weight enumerator, a well-known tool in coding theory \cite{macwilliams1977theory}.
\begin{definition}[Complete Weight Enumerator]
    A complete weight enumerator for a subset $T \subseteq E$ of Pauli errors on $n$ qubits is a homogeneous polynomial $W(i,x,y,z)$ of degree $n$ which is defined as follows:
    \begin{align}
        W(i,x,y,z) = \sum_{M \in T} i^{\mathsf{wt}_I(M)}x^{\mathsf{wt}_X(M)}y^{\mathsf{wt}_Y(M)}z^{\mathsf{wt}_Z(M)}
    \end{align}
    Here $\mathsf{wt}_O(M)$ is the number of occurrences of the Pauli operator $O$ in $M$. In particular, the coefficient of $i^ax^by^cz^d$ is the number of errors in $T$ which have $I$ in $a$ qubits, $X$ in $b$ qubits, $Y$ in $c$ qubits and $Z$ in $d$ qubits.
\end{definition}

\section{Weight Enumerators for Repetition Code}
\label{weightenumerators}
In this section, we describe a closed form expression for coset weight enumerators of repetition codes and concatenated repetition codes. We first describe the enumerators for repetition codes of length $n$. These expressions were also computed in \cite{divincenzo1998quantum}, however they take a different approach by considering the number of non-zero bits in the syndrome measured and then computing the probabilities of corresponding cosets. Our approach has the benefit of generalizing easily to concatenated repetition codes, and the expressions we compute will later also follow as a special case of the concatenated repetition codes. We consider $[[n,1]]$ repetition codes with $Z$-type stabilizers.

\begin{figure}
    \centering
    \includegraphics[width=0.35\linewidth]{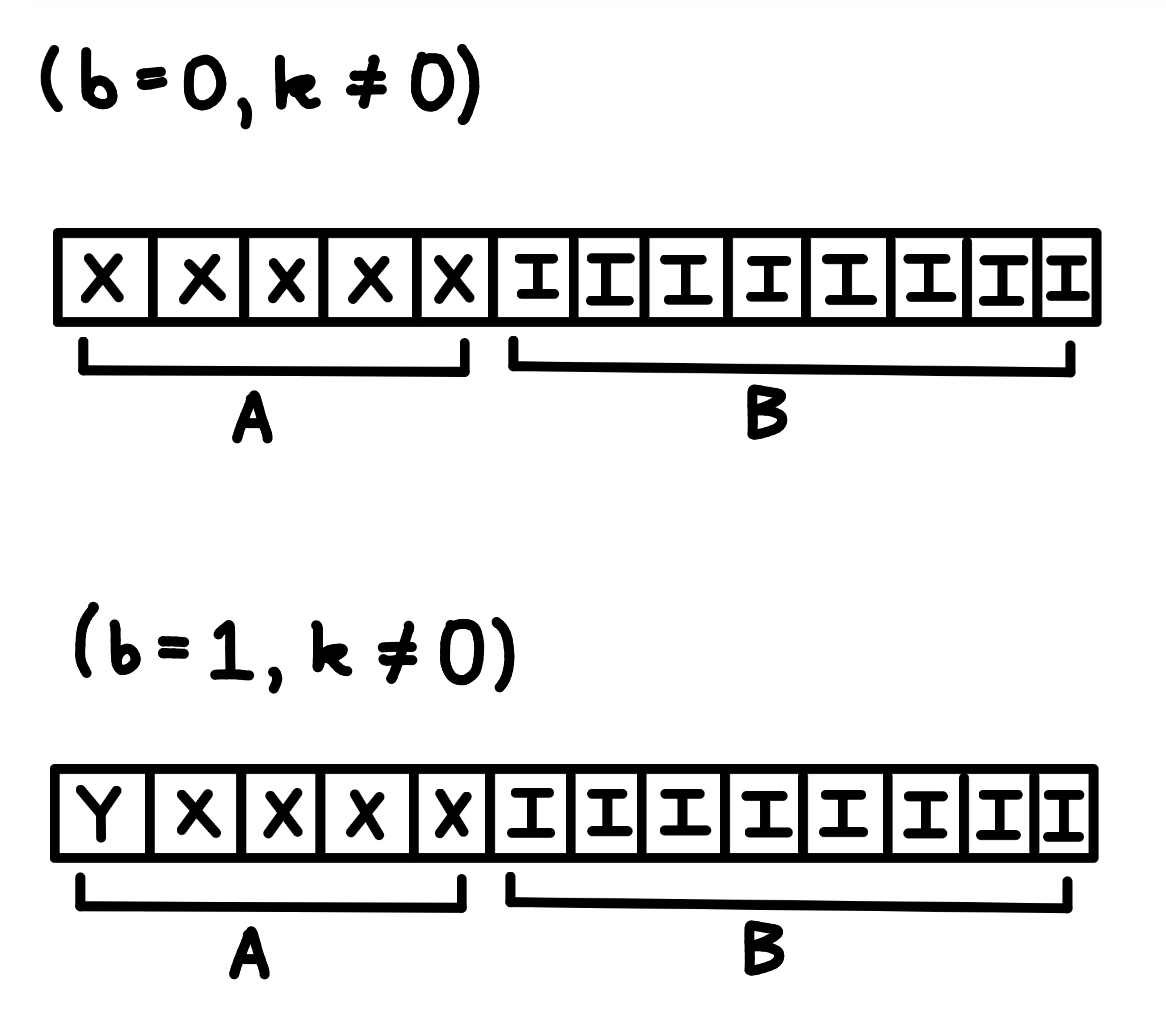}
    \caption{Pictorial representation of Cases 3 and 4}
    \label{fig3}
\end{figure}

The set of stabilizers for the $[[n,1]]$ repetition code is generated by the set $\{Z^{\otimes 2}\otimes I^{\otimes n-2}, I \otimes Z^{\otimes 2} \otimes I^{\otimes n-3}, \ldots, I^{\otimes k}\otimes Z^{\otimes 2} \otimes I^{n-2-k}, \ldots, I^{\otimes n-2} \otimes Z^{\otimes 2}\}$, and therefore is equal to the set of all Pauli strings which satisfy $\mathsf{wt}(Z) = 0 \mod 2$ and $\mathsf{wt}(X) = \mathsf{wt}(Y) = 0$. We choose the representatives $Z \otimes I^{\otimes n-1}, X^{\otimes n}, Y \otimes X^{\otimes n-1}$ for the logical $Z, X, Y$ operations, denoted $\overline{Z}, \overline{X}, \overline{Y}$ respectively. We will also use the following functions to compactly represent our weight enumerators:
\begin{align*}
    f^e_n(x,y) &= \frac{(x+y)^n + (x-y)^n}{2} \\
    f^o_n(x,y) &= \frac{(x+y)^n - (x-y)^n}{2} \\
    g_{k,n}(x,y) &= \frac{1}{2}(x+y)^{n-k}(2y)^{k}
\end{align*}
Here $f^e_n(x,y), f^o_n(x,y)$ are weight enumerators for the set of all even and odd weight binary strings, respectively. The polynomial $g_{k,n}(x,y)$ is a weight enumerator for quaternary strings (say symbols $a,b,c,d$) where $a$ has weight $0$ and the other three have weight $1$. The term $(x+y)^{n-k}$ is a weight enumerator for the set of all strings of $a,b$ on some $n-k$ indices. The term $\frac{1}{2}(2y)^{k}$ is a weight enumerator for the set of all $c,d$ strings which satisfy $\mathsf{wt}(c) = 0 \mod 2$ (or, $\mathsf{wt}(c) = 1 \mod 2$) on some $k$ indices. We can now compute the weight enumerators for the cosets in $E / S$. In the following analysis, we will talk about $\mathsf{wt}_X(M), \mathsf{wt}_Z(M)$ for a Pauli string $M \in E$, where $\mathsf{wt}_X(M)$ is the number of $X$ and $Y$ in $M$ and $\mathsf{wt}_Z(M)$ is the number of $Z$ and $Y$ in $M$, respectively. We point out here that this notation is different from the one used in Section \ref{sec:wtenumrev}, which will not be used here. Note that for any Pauli string $M \in E$, it is equivalent upto stabilizers to a string $M'$ with $\mathsf{wt}_{Z}(M') = \mathsf{wt}_Z(M) \mod 2$, and $\mathsf{wt}_{X}(M') = \mathsf{wt}_X(M)$, where the $Z$-weight is on the first qubit, so now we will only consider coset representatives $M$ of this form. Given a representative string $M$ such that $\mathsf{wt}_X(M) = k$ and $\mathsf{wt}_Z(M) = b$ (where $b \in \{0,1\}$), the weight enumerator for the coset $[M]_S$ will be of the following form:
\begin{itemize}
    \item Case 1: $(b = 0, k = 0)$ This is the case when $M = I^{\otimes n}$ and we want to compute the weight enumerator of the stabilizer, which consists of all $\{I,Z\}$ strings of even $Z$-weight. This results in the expression
    \begin{align*}
        P_{[I^{\otimes n}]_S}(x,y) = f^e_n(x,y).
    \end{align*}
    \item Case 2: $(b = 1, k = 0)$ This is the case when $M = Z \otimes I^{\otimes n-1}$ and we want to compute the weight enumerator of $[M]_S$, which consists of all $\{I,Z\}$ strings of odd $Z$-weight. This results in the expression
    \begin{align*}
        P_{[Z \otimes I^{\otimes n-1}]_S}(x,y) = f^o_n(x,y).
    \end{align*}
    \item Case 3 (See Fig \ref{fig3}): $(b = 0, k \neq 0)$ We want to multiply $M$ with all $\{I,Z\}$ strings of even weight. There are two possibilities, either the stabilizers have odd $Z$-weight in both regions $A$ and $B$ or have even $Z$-weight in both regions. In region $A$, the weight is always $k$ since multiplication of $Z$ converts $X$ to $Y$. This results in the expression
    \begin{align*}
        P_{[M]_S}(x,y) &= f^e_{n-k}(x,y)2^{k-1}y^k + f^o_{n-k}(x,y)2^{k-1}y^k \nonumber \\
        &= \frac{1}{2}(2y)^k(f^e_{n-k}(x,y) + f^o_{n-k}(x,y)) \nonumber \\
        &= \frac{1}{2}(x+y)^{n-k}(2y)^{k} \\
        &= g_{k, n}(x,y).
    \end{align*}
    \item Case 4 (See Fig \ref{fig3}): $(b = 1, k \neq 0)$ We want to multiply $M$ with all $\{I,Z\}$ strings of even weight. Equivalently, we can consider the string $M'$ which is the same string as $M$ but without the $Z$ and multiply it with all $\{I,Z\}$ strings of odd weight. There are two possibilities, either we have odd $Z$-weight in region $A$ and even $Z$-weight in region $B$, or we have even $Z$-weight in region $A$ and odd $Z$-weight in region $B$. In region $A$, the weight is always $k$ since multiplication of $Z$ converts $X$ to $Y$. This results in the expression
    \begin{align*}
        P_{[M]_S}(x,y) &= f^e_{n-k}(x,y)2^{k-1}y^k + f^o_{n-k}(x,y)2^{k-1}y^k \\
        &= \frac{1}{2}(2y)^k(f^e_{n-k}(x,y) + f^o_{n-k}(x,y)) \\
        &= \frac{1}{2}(x+y)^{n-k}(2y)^{k} \\
        &= g_{k, n}(x,y).
    \end{align*}
\end{itemize}
Note that each case above occurs with frequency $\binom{n}{k}$, for the choices of the $k$ possible locations of $X$ in $M$. In order to compute weight enumerators for the normalizer cosets $E/N(S)$, we can add the relevant weight enumerators from the stabilizer cosets. In particular, $[M]_{N(S)} = [M]_S \sqcup [M\overline{Z}]_S \sqcup [M\overline{X}]_S \sqcup [M\overline{Y}]_S$. If $M$ satisfies $\mathsf{wt}_Z(M) = b, \mathsf{wt}_X(M) = k$, then using our choice of representatives for $\overline{Z}, \overline{X}, \overline{Y}$, we can always pick a representative $M_O$ for the coset $[M\overline{O}]_S$ where $O \in \{X, Y, Z\}$ which satisfies the following:
\begin{align*}
    \mathsf{wt}_Z(M_Z) &= 1-b & \mathsf{wt}_X(M_Z) &= k \\
    \mathsf{wt}_Z(M_X) &= b & \mathsf{wt}_X(M_X) &= n-k \\
    \mathsf{wt}_Z(M_Y) &= 1-b & \mathsf{wt}_X(M_Y) &= n-k
\end{align*}
Therefore, the weight enumerator for $[M]_{N(S)}$ can be obtained by adding the weight enumerators for the above cosets:
\begin{align*}
    P_{[M]_{N(S)}}(x,y) = 2(g_{k,n}(x,y) + g_{n-k,n}(x,y))
\end{align*}
where we use the fact that $f^e_n(x,y) + f^o_n(x,y) = 2g_{0,n}(x,y)$ for the case when $k = 0$. We only consider the cases where $b = 0$, since then $\mathsf{wt}_Z(M_Z) = 1$ and $M_Z$ does not have a separate coset. Therefore, this case occurs with frequency $\binom{n}{k}$, for the choices of the $k$ possible locations of $X$ in $M$. Moreover, we now only consider $k \leq \lfloor \frac{n}{2} \rfloor$ since $\mathsf{wt}_X(M_X) = n-k$ so $M_X$ does not have a separate coset. Further, if $n$ is even, we only count half of the $\binom{n}{k}$ choices when $k = n/2$, because for every such $M$ there will be a corresponding $M_X$ with $\mathsf{wt}_X(M_X) = n/2$ which does not have a separate coset. The discussion above can be summarized into the following theorem about the closed form expression for the coset weight enumerators:
\begin{theorem}
    Consider an $[[n,1]]$ repetition code with $Z$-type stabilizers. Let $M = X^{\otimes k} \otimes I^{\otimes (n-k)}$ be a Pauli error on $n$ qubits. The stabilizer and normalizer coset enumerators for $M$ are:
    \begin{align*}
        P_{[M]_S}(x,y) &= \begin{cases}
            f^e_n(x,y) & \text{if } M = I^{\otimes n} \\
            g_{k, n}(x,y) & \text{otherwise}
        \end{cases} \\
        P_{[M\overline{Z}]_S}(x,y) &= \begin{cases}
            f^o_n(x,y) & \text{if } M = I^{\otimes n} \\
            g_{k, n}(x,y) & \text{otherwise}
        \end{cases} \\
        P_{[M\overline{X}]_S}(x,y) &= g_{n-k, n}(x,y) \\
        P_{[M\overline{Y}]_S}(x,y) &= g_{n-k, n}(x,y) \\
        P_{[M]_{N(S)}}(x,y) &= 2(g_{k,n}(x,y) + g_{n-k,n}(x,y))
    \end{align*}
    where $[M]_S$ denotes the stabilizer coset of $M$ and $[M]_{N(S)}$ denotes the normalizer coset of $M$.
\end{theorem}

\subsection{Weight Enumerators for Concatenated Repetition Code}\label{sec:concatwe}
\begin{figure}
    \centering
    \includegraphics[width=0.5\linewidth]{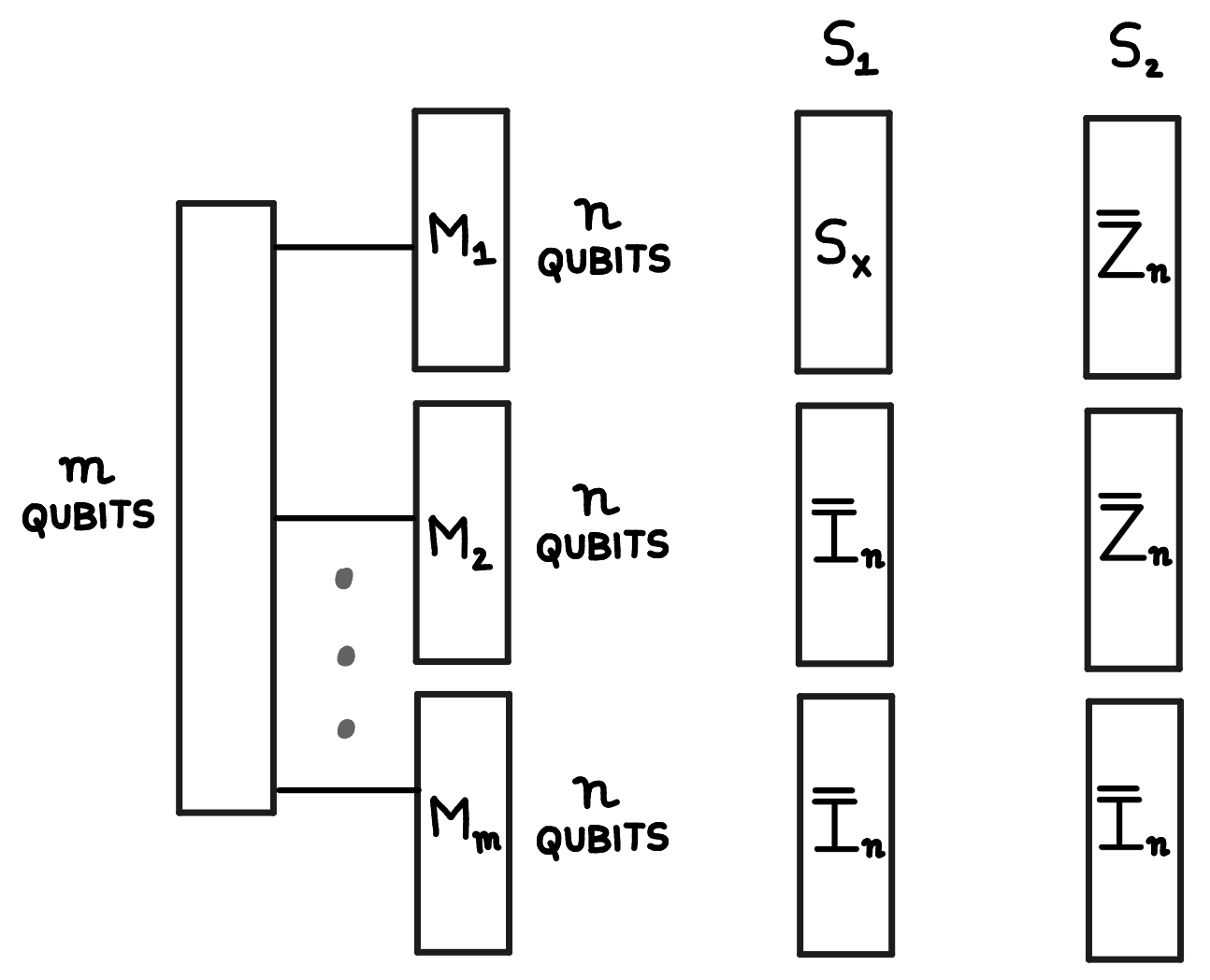}
    \caption{Partitioning of encoded qubits for concatenated repetition codes.}
    \label{fig4}
\end{figure}

In this section, we describe the weight enumerators for the stabilizer cosets of the concatenated repetition code. We will first encode the state using an $[[m,1]]$ repetition code with $Z$-type stabilizer generators $S_Z$ and then encode each of the output qubits using an $[[n,1]]$ repetition code with $X$-type stabilizer generators $S_X$ to get the concatenated $[[mn,1]]$ code $C$. We will think of the $mn$ qubits as $m$ blocks of $n$ qubits each (and the $i^{th}$ block is called $M_i$), see Figure \ref{fig4}. We pick the representatives $\overline{Z}_m = Z \otimes I^{\otimes m-1}, \overline{X}_m = X^{\otimes m}, \overline{Y}_m = Y \otimes X^{\otimes m-1}$ for the logical operations of the $[[m,1]]$ repetition code and $\overline{Z}_n = Z^{\otimes n}, \overline{X}_n = X \otimes I^{\otimes n-1}, \overline{Y}_n = Y \otimes Z^{\otimes n-1}$ for the logical operations of the $[[n,1]]$ repetition code. Given a set $T$ and a Pauli string $P$ we use the notation $T \otimes P$ for the set $T' := \{M \otimes P | M \in T\}$. Given two sets $S_1, S_2 \subseteq E$, we define the set $S_1 \cdot S_2:= \{s_1\cdot s_2| s_i \in S_i\} \subseteq E$.  The resulting set of stabilizers $S$ and representatives of logical operations for $C$ is then:
\begin{align*}
    S_1 &= \langle S_X \otimes \overline{I}_n^{\otimes m-1}, \overline{I}_n \otimes S_X \otimes \overline{I}_n^{\otimes m-2}, \ldots, \overline{I}_n^{\otimes m-1} \otimes S_X \rangle \\
    S_2 &= \langle \overline{Z}_n^{\otimes 2} \otimes \overline{I}_n^{\otimes m-2},\overline{I}_n \otimes \overline{Z}_n^{\otimes 2} \otimes \overline{I}_n^{\otimes m-3}, \ldots, \overline{I}_n^{\otimes m-2} \otimes \overline{Z}_n^{\otimes 2}\rangle \\
    S &= S_1 \cdot S_2 \\
    \overline{Z} &= \overline{Z}_n \otimes \overline{I}_n^{\otimes m-1} \\
    \overline{X} &= (X \otimes I^{\otimes n-1})^{\otimes m} \\
    \overline{Y} &= (Y \otimes Z^{\otimes n-1}) \otimes (X \otimes I^{\otimes n-1})^{\otimes m-1}
\end{align*}
Note here that $S_1$ consists of all strings which have some (potentially different) element of $S_X$ acting in each block of qubits, and $S_2$ consists of strings with $Z^{\otimes n}$ acting on an even number of blocks of qubits and $I^{\otimes n}$ acting on the remaining blocks of qubits. We obtain the set of stabilizers by multiplying each pair of strings picked from $S_1, S_2$. We will use the following functions to compactly represent our weight enumerators:
\begin{align*}
    F_{n}^{e}(\vec{x},\vec{y}) &= \frac{1}{2}\left[\prod_{i=1}^{n}(x_i+y_i)+\prod_{i=1}^{n}(x_i-y_i)\right] \\
    F_{n}^{o}(\vec{x},\vec{y}) &= \frac{1}{2}\left[\prod_{i=1}^{n}(x_i+y_i)-\prod_{i=1}^{n}(x_i-y_i)\right] \\
    h_{k,n}^{b}(x,y) &= \begin{cases}
f_{n}^{e}(x,y) & \text{if } k=0, b=0 \\
f_{n}^{o}(x,y) & \text{if } k=0, b=1 \\
g_{k,n}(x,y) & \text{otherwise}
\end{cases}
\end{align*}
Here $F_{n}^{e}(\vec{x},\vec{y}), F_{n}^{o}(\vec{x},\vec{y})$ take vector-valued variables $\vec{x}, \vec{y}$ as input each of length $n$. The polynomial $F_{n}^{e}(\vec{x},\vec{y})~(F_{n}^{o}(\vec{x},\vec{y}))$ consists of all possible choices of $x_i$ or $y_i$ for each $i$, where an even (odd) number of variables are chosen from $\vec{y}$. The polynomial $h_{k,n}^{b}(x,y)$ computes the weight distribution for the different cases from the case of non-concatenated repetition codes. We can now compute the weight enumerators for the cosets in $E / S$. In the following analysis, we will talk about $\mathsf{wt}_{\vec{X}}(M), \mathsf{wt}_{\vec{Z}}(M)$ for a Pauli string $M \in E$, where we will divide $M$ into $m$ blocks of length $n$, denote the $i^{th}$ block by $M_i$, and count the Pauli weights for each block separately. Then $\mathsf{wt}_{\vec{X}}(M), \mathsf{wt}_{\vec{Z}}(M)$ are vectors of length $m$, where the $i^{th}$ entry of $\mathsf{wt}_{\vec{X}}(M)$ ($\mathsf{wt}_{\vec{Z}}(M)$) is the number of $X$ and $Y$ ($Z$ and $Y$) in the $i^{th}$ block of $M$. Note that for any $M \in E$, it is equivalent up to stabilizers to a string $M'$ with $\mathsf{wt}_{\vec{Z}}(M') = \mathsf{wt}_{\vec{Z}}(M)$ and $\mathsf{wt}_{\vec{X}}(M') = \mathsf{wt}_{\vec{X}}(M) \mod 2$ by multiplying with the stabilizer elements of $S_1$, and the $X$ weight in each block of $M$ is on the first qubit of the block. Now we will only consider $M$ of this form, which has $\mathsf{wt}_{\vec{X}}(M) = \vec{b}, \mathsf{wt}_{\vec{Z}}(M) = \vec{k}$ where $\vec{b} \in \{0,1\}^m, \vec{k} \in [n]^m$. Then we can compute the weight enumerators of $[M]_S, [M\overline{Z}]_S, [M\overline{X}]_S, [M\overline{Y}]_S$ as follows:
\begin{itemize}
    \item For $[M]_S$: We want to compute a weight enumerator for the set of all strings generated by multiplying $M$ with pairs of strings from $S_1, S_2$. We will first focus on the weight distribution in the first block $(M_1)$. Suppose we pick a string from $S_2$ that has $I^{\otimes n}$ in the first block. Then we want to compute the weight distribution of the strings obtained by multiplying $M_1$ with all the stabilizer strings in $S_X$, since $S_2$ consists of strings with an element of $S_X$ in each block. In particular, we want to compute the weight distribution of strings obtained by multiplying $M_1$ with all $\{I,X\}$ strings of even weight. Depending on the value of $b_1$ and $k_1$, this will be the same as one of the cases we considered from non-concatenated repetition codes. Thus this expression will be $h^{b_1}_{k_1, n}(x,y)$. If we picked a string from $S_2$ that has $Z^{\otimes n}$ in the first block then we can equivalently consider a string $M_1'$ which has $\mathsf{wt}_Z(M_1') = n-k_1$, $\mathsf{wt}_X(M_1') = b_1$ and then compute the weight distribution of strings obtained by multiplying $M_1'$ with all $\{I,X\}$ strings of even weight. Thus this expression will be $h^{b_1}_{n-k_1, n}(x,y)$. Now to compute the full weight distribution, we want to consider all the blocks of $M$. In particular, on multiplying by elements in $S_2$, there will be an even number of blocks which have $Z^{\otimes n}$ multiplied with them. These blocks will have the weight distribution $h^{b_i}_{n-k_i, n}(x,y)$ and the rest will have the weight distribution $h^{b_i}_{k_i, n}(x,y)$. So we will consider a vector valued function $\vec{h}^{\vec{b}}_{\vec{k},\vec{n}}(x,y)$ of length $m$ where the $i^{th}$ coordinate is $h_{k_i,n}^{b_i}(x,y)$. Then the overall weight distribution is
    \begin{align*}
        P_{[M]_S}(x,y) = F_{m}^{e}(\vec{h}^{\vec{b}}_{\vec{k},\vec{n}}(x,y),\vec{h}^{\vec{b}}_{\vec{n}-\vec{k},\vec{n}}(x,y)).
    \end{align*}
    \item For $[M\overline{Z}]_S$: Note that $\overline{Z} = \overline{Z}_n \otimes \overline{I}_n^{\otimes m-1}$. To compute the weight distribution of $[M\overline{Z}]_S$, we want to multiply $M\overline{Z}$ with pairs of strings from $S_1, S_2$. Note that every string in $S_2$ on being multiplied with $\overline{Z}$ becomes a string which has $Z^{\otimes n}$ in an odd number of blocks and $I^{\otimes n}$ in an even number of blocks. So we can do a computation similar to $[M]_S$ but with an odd number of blocks having $Z^{\otimes n}$ multiplied with them. These blocks will have the weight distribution $h^{b_i}_{n-k_i, n}(x,y)$ and the rest will have the weight distribution $h^{b_i}_{k_i, n}(x,y)$. Then the overall weight distribution is
    \begin{align*}
        P_{[M\overline{Z}]_S}(x,y) = F_{m}^{o}(\vec{h}^{\vec{b}}_{\vec{k},\vec{n}}(x,y),\vec{h}^{\vec{b}}_{\vec{n}-\vec{k},\vec{n}}(x,y)).
    \end{align*}
    \item For $[M\overline{X}]_S$: Note that $\overline{X} = (X \otimes I^{\otimes n-1})^{\otimes m}$. To compute the weight distribution of $[M\overline{X}]_S$, we want to multiply $M\overline{X}$ with pairs of strings from $S_1, S_2$. Since multiplication with $\overline{X}$ changes the parity of every entry in $\mathsf{wt}_{\vec{X}}(M)$, we can choose a representative $M'$ of $[M\overline{X}]_S$ which has $\mathsf{wt}_{\vec{X}}(M') = (\mathsf{wt}_{\vec{X}}(M))^c$ (where $\vec{v}^c$ denotes entry-wise complement) and $\mathsf{wt}_{\vec{Z}}(M') = \mathsf{wt}_{\vec{Z}}(M)$. Then we can do a computation similar to $[M]_S$ where we want to multiply $M'$ with pairs of string from $S_1, S_2$. Then the overall weight distribution is
    \begin{align*}
        P_{[M\overline{X}]_S}(x,y) = F_{m}^{e}(\vec{h}^{\vec{b}^c}_{\vec{k},\vec{n}}(x,y),\vec{h}^{\vec{b}^c}_{\vec{n}-\vec{k},\vec{n}}(x,y)).
    \end{align*}
    \item For $[M\overline{Y}]_S$: Note that $\overline{Y} = (Y \otimes Z^{\otimes n-1}) \otimes (X \otimes I^{\otimes n-1})^{\otimes m-1} = (\overline{Z}_n \otimes \overline{I}_n^{\otimes m-1})((X \otimes I^{\otimes n-1})^{\otimes m})$. To compute the weight distribution of $[M\overline{Y}]_S$, we can now pick a representative $M'$ of $[M\overline{Y}]_S$ which has $\mathsf{wt}_{\vec{X}}(M') = (\mathsf{wt}_{\vec{X}}(M))^c$ and $\mathsf{wt}_{\vec{Z}}(M') = \mathsf{wt}_{\vec{Z}}(M\overline{Z})$. In addition, multiplying $(\overline{Z}_n \otimes \overline{I}_n^{\otimes m-1})$ with strings in $S_2$ converts each string to one having odd number of blocks having $Z^{\otimes n}$. These blocks will have the weight distribution $h^{b_i}_{n-k_i, n}(x,y)$ and the rest will have the weight distribution $h^{b_i}_{k_i, n}(x,y)$. Then the overall weight distribution is
    \begin{align*}
        P_{[M\overline{Y}]_S}(x,y) = F_{m}^{o}(\vec{h}^{\vec{b}^c}_{\vec{k},\vec{n}}(x,y),\vec{h}^{\vec{b}^c}_{\vec{n}-\vec{k},\vec{n}}(x,y)).
    \end{align*}
\end{itemize}
To compute the weight enumerators for the normalizer cosets $E/N(S)$, we can add the relevant weight enumerators from the stabilizer cosets. In particular, to compute $P([M]_{N(S)})$:
\begin{align*}
    P_{[M]_{N(S)}}(x,y) &= P_{[M]_S}(x,y) + P_{[M\overline{Z}]_S}(x,y) + P_{[M\overline{X}]_S}(x,y) + P_{[M\overline{Y}]_S}(x,y) \nonumber \\
    &= \left(\prod_{i=1}^m(h^{b_i}_{k_i,n}(x,y) +h^{b_i}_{n-k_i,n}(x,y))\right) + \left(\prod_{i=1}^m(h^{1-b_i}_{k_i,n}(x,y) +h^{1-b_i}_{n-k_i,n}(x,y))\right)
\end{align*}
where we have used the fact that $F_{n}^{e}(\vec{x},\vec{y}) + F_{n}^{o}(\vec{x},\vec{y}) = \prod_{i=1}^n(x_i+y_i)$. The discussion above can be summarized in the following theorem about the closed form expression for the coset weight enumerators:

\begin{theorem}
    Consider an $n \times m$ concatenated bit and phase-flip repetition code. Let $M \in E$ be a Pauli error on $mn$ qubits. The stabilizer and normalizer coset weight enumerators for $M$ are:
    \begin{align*}
        P_{[M]_S}(x,y) &= F_{m}^{e}(\vec{h}^{\vec{b}}_{\vec{k},\vec{n}}(x,y),\vec{h}^{\vec{b}}_{\vec{n}-\vec{k},\vec{n}}(x,y)) \\
        P_{[M\overline{Z}]_S}(x,y) &= F_{m}^{o}(\vec{h}^{\vec{b}}_{\vec{k},\vec{n}}(x,y),\vec{h}^{\vec{b}}_{\vec{n}-\vec{k},\vec{n}}(x,y)) \\
        P_{[M\overline{X}]_S}(x,y) &= F_{m}^{e}(\vec{h}^{\vec{b}^c}_{\vec{k},\vec{n}}(x,y),\vec{h}^{\vec{b}^c}_{\vec{n}-\vec{k},\vec{n}}(x,y)) \\
        P_{[M\overline{Y}]_S}(x,y) &= F_{m}^{o}(\vec{h}^{\vec{b}^c}_{\vec{k},\vec{n}}(x,y),\vec{h}^{\vec{b}^c}_{\vec{n}-\vec{k},\vec{n}}(x,y))
    \end{align*}
    \begin{align*}
        P_{[M]_{N(S)}}(x,y) &= \left(\prod_{i=1}^m(h^{b_i}_{k_i,n}(x,y) +h^{b_i}_{n-k_i,n}(x,y))\right) + \left(\prod_{i=1}^m(h^{1-b_i}_{k_i,n}(x,y) +h^{1-b_i}_{n-k_i,n}(x,y))\right)
    \end{align*}
    where $[M]_S$ denotes the stabilizer coset of $M$, $[M]_{N(S)}$ denotes the normalizer coset of $M$ and $\vec{b}, \vec{k}$ are as defined in the discussion above.

\end{theorem}

\subsection{Description of threshold estimation algorithm}
\label{algorithm}
\paragraph{Overview:}

Recall that the entropy $S_{RB}$ can be computed from the coset weight enumerators as
\begin{align*}\label{eq:ent}
\sum_{T \in E/N(S)}P_{T}(1-3p, p) H\left(\frac{P_{T_{\overline{I}}}(1-3p, p)}{P_{T}(1-3p, p)}, \frac{P_{T_{\overline{X}}}(1-3p, p)}{P_{T}(1-3p, p)}, \frac{P_{T_{\overline{Y}}}(1-3p, p)}{P_{T}(1-3p, p)}, \frac{P_{T_{\overline{Z}}}(1-3p, p)}{P_{T}(1-3p, p)}\right).
\end{align*}

This expression can rewritten as the difference between the entropies attributed to the stabilizer coset and the normalizer coset.
\begin{align*}
\underbrace{\sum_{T \in E/S} -P_T(1 - 3p, p) \ln P_T(1 - 3p, p)}_{\textrm{entropy of stabilizer coset}}- \underbrace{\sum_{T \in E/N(S)} -P_T(1 - 3p, p)\ln P_T(1 - 3p, p)}_{\textrm{entropy of normalizer coset}}
\end{align*}
In this section, we describe an algorithm that estimates each half of this expression for concatenated repetition codes by using the weight enumerators from the previous section. This enables us to efficiently analyze $S_{RB}$ as a function of $p$ and find the error threshold $p_*$ when $S_{RB} = 1$.

\paragraph{Stabilizer coset entropy:}
Using the concatenated weight enumerator expressions from Section \ref{sec:concatwe}, the probability of observing the stabilizer coset $[M]_S$ is given by
\[P_{[M]_S} = \frac{1}{2}\left[ \prod_{i = 1}^m (h_{k_i, n}^{b_i} + h_{n-k_i, n}^{b_i}) + \prod_{i = 1}^m (h_{k_i, n}^{b_i} - h_{n-k_i, n}^{b_i})\right],\]

so the entropy attributed to the stabilizer coset can be calculated as
\begin{align*}
&\quad \sum_{[M]_S \in E/S} -P_{[M]_S}(1 - 3p, p) \ln P_{[M]_S}(1 - 3p, p) = \mathbb{E}[-\ln P_{[M]_S}(1-3p, p)] \\
&= \ln 2 + n\,\mathbb{E}[-\ln(h_{k, n}^b + h_{n-k, n}^b)] + \mathbb{E}\left[-\ln\left(1 + \prod_{i = 1}^m \left( \frac{h_{k_i, n}^{b_i} - h_{n-k_i, n}^{b_i}}{h_{k_i, n}^{b_i} + h_{n-k_i, n}^{b_i}}\right)\right)\right],
\end{align*}

where the choices of pairs $(k_i, b_i)_{i \in [n]}$ are mutually independent. The term $\mathbb{E}[- \ln(h_{k, n}^b + h_{n-k, n}^b)]$ is independent of $m$ and can be computed in $O(n)$ time. For each pair $(k, b) \in [n] \times \{0, 1\},$ we define the coefficient
\[q_{k, n}^b = \frac{h_{k, n}^b - h_{n-k, n}^b}{h_{k, n}^b + h_{n-k, n}^b} \in (-1, 1).\]

The last term is re-expressed in a form efficiently computed by the algorithm
\[\mathbb{E}\left[-\ln\left(1 + \prod_{i = 1}^m q_i\right)\right], \textrm{ each }q_i \sim Q,\]

where $Q$ is the single-channel distribution of $q_{k, n}^b$. The algorithm then convolves the distribution of $(\textrm{sign}(q), \ln |q|) \in \mathbb{Z}/2 \times \mathbb{R}$ derived from $q \sim Q$ with itself $n$ times, and evaluates the expectation of $-\ln(1 \pm e^{x})$ on the resulting distribution, where the choice of $\pm$ is given by the overall sign of $\prod q_i$ and $x$ is the sum of $\ln |q_i|$. This yields the entropy attributed to the stabilizer coset.

\paragraph{Normalizer coset entropy:}

From Section \ref{sec:concatwe}, the probability of observing the normalizer coset $[M]_{N(S)}$ is given by
\[P_{[M]_{N(S)}} = \prod_{i = 1}^m(h_{k_i, n}^{b_i} + h_{n-k_i, n}^{b_i}) + \prod_{i = 1}^m(h_{k_i, n}^{1-b_i} + h_{n-k_i, n}^{1-b_i}).\]

Thus, the entropy attributed to the normalizer coset can be calculated as
\begin{align*}
    &\quad \sum_{[M]_{N(S)} \in E/N(S)}-P_{[M]_{N(S)}}(1 - 3p, p) \ln P_{[M]_{N(S)}}(1 - 3p, p) = \mathbb{E}[-\ln P_{[M]_{N(S)}}(1 - 3p, p)] \\
    &= n\,\mathbb{E}[-\ln(h_{k, n}^b + h_{n-k, n}^b)] + \mathbb{E}\left[ -\ln\left( 1 + \prod_{i = 1}^m \left( \frac{h_{k, n}^{1-b} + h_{n-k, n}^{1-b}}{h_{k, n}^b + h_{n-k, n}^b} \right)\right)\right],
\end{align*}

where the choices of pairs $(k_i, b_i)_{i \in [n]}$ are mutually independent. The first term can be reused. For each pair $(k, b) \in [n] \times \{0, 1\},$ we define the coefficient
\[r_{k, n}^b = \frac{h_{k, n}^{1-b} + h_{n-k, n}^{1-b}}{h_{k, n}^b + h_{n-k, n}^b} \in \mathbb{R}_+.\]

The relevant term is re-expressed in a form efficiently computed by the algorithm
\[\mathbb{E}\left[-\ln\left( 1 + \prod_{i = 1}^m r_i\right)\right], \textrm{ each }r_i \sim R,\]

where $R$ is the single-channel distribution of $r_{k, n}^b$. The algorithm then convolves the distribution $\ln R$ with itself $n$ times, and evaluates the expectation of $-\ln(1 + e^x)$ on the resulting distribution. This yields the entropy attributed to the normalizer coset.

To perform convolutions over $\mathbb{R},$ we bin the distributions $Q$ and $R$ by rounding them to $\alpha \cdot \mathbb{Z}$ for a scale factor $\alpha \in \mathbb{R}_+$, and perform a fast Fourier transform on the discretized data. We repeat this bin-and-convolve process at $O(n)$ different scales, and recombine the post-convolution data. The binning procedure and choice of scales is discussed in the GitHub repository \cite{github}.\footnote{See the folder ``Threshold Estimation"}

Note that the algorithm becomes unstable for very long lengths in the second layer, and we only report stable values for different lengths.

\section{Search over Stabilizer Codes}
\subsection{Depolarizing Channel}
In this section, we study the problem for computing the threshold for the depolarizing channel. For completeness, we also report some values that have been reported in prior works (in particular, \cite{fern}). 
\begin{table}
\centering
\begin{minipage}{0.48\textwidth}
\centering
\begin{tabular}{|l|>{\ttfamily}p{3.0cm}|}
\hline
\textbf{Code} & \textbf{Threshold} \\
\hline
5 rep & 0.06345202939 \\
\hline
7 rep & 0.06341083749 \\
\hline
3 rep & 0.06337664297 \\
\hline
Shor code (9-qubit) & 0.06335987828 \\
\hline
4 rep & 0.06329834885 \\
\hline
\color{red}{Hashing} & \red{0.0630965616} \\
\hline
5 qubit & 0.06298730942 \\
\hline
Biased 13-qubit & 0.0628392531 \\
\hline
$[[11, 1, 5]]$ & 0.06283811205 \\
\hline
\end{tabular}
\end{minipage}
\hfill
\begin{minipage}{0.48\textwidth}
\centering
\begin{tabular}{|l|>{\ttfamily}p{3.0cm}|}
\hline
\textbf{Code} & \textbf{Threshold} \\
\hline
13-qubit cyclic code & 0.06277288597 \\
\hline
Tailored $[[7, 1, 3]]$ H & 0.06276470396 \\
\hline
$[[6, 1, 3]]$ H & 0.0627509475 \\
\hline
Biased 9-qubit & 0.06275087308 \\
\hline
$[[4,2,2]]$ & 0.06261572 \\
\hline
7 qubit (Steane) & 0.06259214551 \\
\hline
CD Steane H & 0.06259214551 \\
\hline
Toric code $[[8, 2, 2]]$ & 0.06247322092 \\
\hline
SCF H & 0.06236285581 \\
\hline
\end{tabular}
\end{minipage}
\caption{Error Thresholds for various stabilizer codes (depolarizing channel). The codes with an H refer to holographic codes, discussed in \cite{fan2024overcoming}. SCF is Surface Code Fragment \cite{scf}. CD Steane is Clifford deformed Steane code. The [[6,1,3]] and tailored [[7,1,3]] codes were proposed in \cite{6qubit, tailored7} respectively.}
\label{table1}
\end{table}
We start by considering a single layer of various stabilizer codes. The thresholds for these codes are reported in Table \ref{table1}. The only codes which give a better threshold than hashing are repetition codes or the Shor code (which is a concatenation of 2 repetition codes). In particular, note that even codes which have non-trivial distance (e.g. 5-qubit code) also do worse than random hashing. This is in contrast with the situation in quantum error correction wherein codes such as the 5-qubit code, the Steane code and the toric code perform well in different scenarios. For example, the 5-qubit code has the best threshold for magic state distillation for $T$-states \cite{rall2017signed, kalra2025invariant}.

Note that repetition codes are classical codes (stabilizers of a single type X or Z) so intuitively one would expect that they only work well for channels with a single type of error. A possible explanation for the good performance of repetition codes has been suggested in \cite{shor1996quantumerrorcorrectingcodesneed, smith2007degenerate}, who posit that repetition codes are degenerate for Z errors, and they can correct X errors, thus helping reduce the overall entropy of the state. However, on the other hand, codes such as the biased 9 and 13 qubit code that perform well for biased channels \cite{biased} (but are not completely classical) have a significantly worse threshold. 

\begin{table}
\centering
\small
\begin{minipage}{0.48\textwidth}
\centering
\begin{tabular}{|l|>{\ttfamily}p{3.0cm}|}
\hline
\textbf{Code} & \textbf{Threshold} \\
\hline
5 rep Z $\times$ 5 rep X & 0.06352047429 \\
\hline
5 rep Z $\times$ Biased 9-qubit & 0.063514550053 \\
\hline
5 rep $\times$ 5 qubit & 0.06350293118 \\
\hline
7 rep Z $\times$ Biased 9-qubit (E) & 0.063480612253 \\
\hline
7 rep $\times$ 5 qubit & 0.06345748411 \\
\hline
\red{5 rep} & \red{0.06345202939} \\
\hline
7 rep & 0.06341083749 \\
\hline
5 rep X $\times$ Biased 9-qubit & 0.063409518776 \\
\hline
3 rep & 0.06337664297 \\
\hline
3 rep $\times$ 7 qubit & 0.06336876052 \\
\hline
3 rep $\times$ Toric code [[8,2,2]] & 0.06335777052 \\
\hline
3 rep $\times$ Biased 13-qubit (E) & 0.06334324161  \\
\hline
3 rep Z $\times$ Biased 9-qubit & 0.063340626283 \\
\hline
\end{tabular}
\end{minipage}
\hfill
\begin{minipage}{0.48\textwidth}
\centering
\begin{tabular}{|l|>{\ttfamily}p{3.0cm}|}
\hline
\textbf{Code} & \textbf{Threshold} \\
\hline
4 rep $\times$ [[4,2,2]] & 0.06333945247 \\
\hline
3 rep $\times$ 5 qubit & 0.06333924742 \\
\hline
3 rep X $\times$ 13-qubit code (E) & 0.06332969723 \\
\hline
4 rep $\times$ Toric code [[8,2,2]] & 0.0633281095 \\
\hline
3 rep $\times$ [[4,2,2]] & 0.06332006618 \\
\hline
4 rep $\times$ 5 qubit & 0.06330111376 \\
\hline
5 rep $\times$ [[4,2,2]] & 0.06329169236 \\
\hline
3 rep X $\times$ Biased 9-qubit & 0.063291502219 \\
\hline
5 rep $\times$ Toric code [[8,2,2]] & 0.06324287112 \\
\hline
5 rep $\times$ 7 qubit & 0.06310170543 \\
\hline
\red{Hashing} & \red{0.0630965616} \\
\hline
7 rep $\times$ 7 qubit & 0.0627916763 \\
\hline
\end{tabular}
\end{minipage}
\caption{Error Threshold for repetition codes concatenated with small stabilizer codes (depolarizing channel). Concatenating 5-repetition with biased 9-qubit improves the threshold. Entries with an (E) refer to Monte-Carlo based estimates.}
\label{table2}
\end{table}
We then study the error thresholds for repetition codes concatenated with several small stabilizer codes. The thresholds are reported in Table \ref{table2}. One question we consider is whether any small concatenations can outperform the 5-repetition code (best for single-layer). We start by concatenating repetition codes with the biased 9-qubit code, to see if it improves the threshold. The biased 9-qubit code designed in \cite{biased} is used to correct one generic error and one extra Z error. We perform such a concatenation  motivated by the observation that the output channels of repetition codes with Z stabilizers are significantly biased towards Z errors, and hence concatenation with the biased 9-qubit should help improve the threshold. We find this is indeed the case for the 5-repetition and 7-repetition code, but not for the 3-repetition code. One reason for not seeing the improvement at 3-rep could be that 5-rep and 7-rep codes result in more biased effective channels. Note also that even though the 7-rep case shows an improvement after concatenation, the threshold is worse than that of the 5-rep case concatenated with biased 9-qubit. Note that concatenations of this type have not been previously studied. 

Concatenation of repetition codes with the 5-qubit code improves the threshold in all cases except 3-repetition code. This pattern is not seen for the 7-qubit code, which always makes the threshold worse in such a concatenation. Concatenation with the [[8,2,2]] toric code and [[4,2,2]] code improves the threshold for 4-repetition code, but not for odd length repetition codes. However, the improvement is not enough to make it do better than 5-repetition. When concatenating repetition codes with the toric code, increasing the length of the repetition code makes the threshold worse.
\begin{table}
\centering
\footnotesize
\begin{minipage}[t]{0.48\textwidth}
\centering
\begin{tabular}{|l|>{\ttfamily}p{3.0cm}|}
\hline
\textbf{Concatenated Codes} & \textbf{Value} \\
\hline
5 rep $\times$ Tailored [[7,1,3]] H & 0.063513617158 \\
\hline
5 rep $\times$ 5 qubit & 0.06350293118 \\
\hline
5 rep X $\times$ [[6,1,3]] H & 0.063485121221 \\
\hline
5 rep $\times$ CD Steane H & 0.06347779536 \\
\hline
7 rep $\times$ Tailored [[7,1,3]] H & 0.06347148479 \\
\hline
7 rep $\times$ 5 qubit & 0.06345748411 \\
\hline
\red{5 rep} & \red{0.06345202939} \\
\hline
7 rep X $\times$ [[6,1,3]] H & 0.063446922 \\
\hline
7 rep $\times$ CD Steane H & 0.06343681225 \\
\hline
7 rep & 0.06341083749 \\
\hline
3 rep & 0.06337664297 \\
\hline
3 rep $\times$ SCF H & 0.06334961171 \\
\hline
4 rep $\times$ [[6,1,3]] H & 0.06334802787 \\
\hline
4 rep $\times$ SCF H & 0.06333589217 \\
\hline
Shor code $\times$ [[6,1,3]] H & 0.06333181819 \\
\hline
3 rep $\times$ Tailored [[7,1,3]] H & 0.063331611435 \\
\hline
\end{tabular}
\end{minipage}
\hfill
\begin{minipage}[t]{0.48\textwidth}
\centering
\begin{tabular}{|l|>{\ttfamily}p{3.0cm}|}
\hline
\textbf{Concatenated Codes} & \textbf{Value} \\
\hline
3 rep X $\times$ [[6,1,3]] H & 0.063314378012 \\
\hline
4 rep $\times$ CD Steane H & 0.06331188169 \\
\hline
5 rep Z $\times$ [[6,1,3]] H & 0.06330793247 \\
\hline
4 rep $\times$ Tailored [[7,1,3]] H & 0.06330435597 \\
\hline
3 rep Z $\times$ [[6,1,3]] H & 0.063299275357 \\
\hline
3 rep $\times$ CD Steane H & 0.06329370446 \\
\hline
5 rep $\times$ SCF H & 0.06325328184 \\
\hline
7 rep Z $\times$ [[6,1,3]] H & 0.063124747103 \\
\hline
\red{Hashing} & \red{0.0630965616} \\
\hline
7 rep $\times$ SCF H & 0.06305013433 \\
\hline
5-qubit & 0.06298730942 \\
\hline
5-qubit $\times$ [[6,1,3]] H & 0.06298013551 \\
\hline
5-qubit $\times$ SCF H & 0.06296798167 \\
\hline
7-qubit $\times$ SCF H & 0.06266241974 \\
\hline
7-qubit $\times$ [[6,1,3]] H & 0.06266195636 \\
\hline
$[[4,2,2]]$ $\times$ [[6,1,3]] H & 0.0625556411 \\
\hline
\end{tabular}
\end{minipage}
\vspace{0.3cm}
\caption{Error Thresholds for stabilizer codes concatenated with holographic codes (depolarizing channel). Multiple concatenations lead to improved thresholds.}
\label{table3}
\end{table}

The next set of numerical experiments we carried out were to study the thresholds of repetition codes concatenated with holographic codes. The thresholds are reported in Table \ref{table3}. This experiment was inspired by recent results in \cite{fan2024overcoming}, where the authors show that holographic codes perform better than hashing for biased noise channels. We find that on concatenating repetition codes with holographic codes, several of these codes do better than repetition codes alone. For small code concatenations, these and the biased 9-qubit code are the first examples of stabilizer codes that are not repetition codes which improve the threshold when used in the second layer.

For 5-repetition, concatenation with tailored [[7,1,3]] code, [[6,1,3]] code and the CD Steane code, all improve the threshold. All these codes also show improvement in the threshold for 7-repetition code. The best threshold is obtained by concatenating 5-repetition with the tailored [[7,1,3]] code. For 4-repetition code, concatenating it with the [[6,1,3]] code and the SCF code improves the threshold, and the best threshold is given by the [[6,1,3]] code. Concatenating a non-repetition code such as 5-qubit, 7-qubit or [[4,2,2]] code with a holographic code always results in a threshold worse than hashing. We note that none of these concatenations have been previously studied.
\begin{table}
\centering
\small
\begin{tabular}{|l|>{\ttfamily}p{3.0cm}|}
\hline
\textbf{Code} & \textbf{Threshold} \\
\hline
5 rep X $\times$ 5 qubit $\times$ 5 rep Z & 0.063552 (E) \\ \hline
5 rep $\times$ 7 perm & 0.06354354568 \\ \hline
5 rep X $\times$ 5 rep Z $\times$ 5 qubit & 0.063535 (E)\\ \hline
5 rep $\times$ 5 perm & 0.06352047429 \\ \hline
5 rep X $\times$ 5 qubit $\times$ 5 qubit & 0.063519 (E) \\ \hline
5 rep Z $\times$ Biased 9-qubit & 0.063514550053 \\ \hline
5 rep $\times$ 3 perm & 0.06349265479 \\ \hline
3 rep $\times$ 7 perm & 0.0634785941 \\ \hline
\red{5 rep} & \red{0.06345202939} \\ \hline
3 rep $\times$ 5 perm & 0.06341736499 \\ \hline
5 rep X $\times$ Biased 9-qubit & 0.063409518776 \\ \hline
3 rep Z $\times$ Biased 9-qubit & 0.063340626283 \\ \hline
3 rep X $\times$ Biased 9-qubit & 0.063291502219 \\ \hline
5 rep X $\times$ 5 rep Z $\times$ 5 rep X & 0.063278 (E) \\ \hline
3 rep X $\times$ 3 rep Z $\times$ 3 rep X & 0.06327013619 \\ \hline
\red{Hashing} & \red{0.0630965616} \\ \hline
\end{tabular}
\caption{A collection of some conceptually interesting codes (depolarizing channel). Entries with an (E) refer to Monte-Carlo based estimates. Perm refers to permutation-invariant code, and we find that the best permutation-invariant codes are repetition codes for our examples.}
\label{table4}
\end{table}

We then study the thresholds for a number of codes which are conceptually interesting and have not been explored before. The thresholds are reported in Table \ref{table4}. In these experiments, we first study the threshold for 3 layers of repetition codes, for length 3 and 5 repetition codes. One reason to study these concatenations is that two layers of repetition codes lead to better thresholds than a single layer, so it is natural to wonder whether further concatenations improve the threshold. In both cases, the threshold is worse than that obtained using a single layer of repetition code. Another reason for studying these 3 layers of repetition code is that the output channel after two layers of repetition code is highly biased towards one error, in which case concatenating with another layer of repetition code should help correct that error, thus reducing entropy of the overall state. However, the result is in contrast with previous results (Tables \ref{table2} and \ref{table3}) where concatenation with codes for biased channels improves the threshold. One reason for this could be that repetition codes in the third layer only have stabilizers of one type, while the other codes have both X and Z stabilizers.

We next study the thresholds for the following combination of codes: 2 layers of repetition codes and one layer of 5-qubit code. A combination of these codes was used in \cite{fern} to achieve the best thresholds (though for much longer lengths). We arrange these codes in different configurations to study which configuration yields the best threshold: 5 rep X $\times$ 5 qubit $\times$ 5 rep Z and 5 rep X $\times$ 5 rep Z $\times$ 5 qubit. We see that the first configuration has the better threshold and both configurations do better than two layers of 5-repetition codes. We note that previous works have not studied any concatenations with 5-qubit code in the middle layer. Furthermore, we see that concatenating the 5-repetition code with two layers of 5-qubit code gives a better threshold than concatenating with a single layer of 5-qubit code, which in turn gives a better threshold than just the 5-repetition code. Finally, motivated by the recent results of \cite{bhalerao2025improvingquantumcommunicationrates} we concatenate repetition codes with permutation invariant codes, and optimize over the choice of permutation invariant codes for small lengths. Similar to the results of \cite{bhalerao2025improvingquantumcommunicationrates} for the depolarizing channel, we find that even after concatenating with the repetition code in first layer, the best permutation invariant code remains the repetition code.
\begin{table}
\centering
\begin{tabular}{|l|>{\ttfamily}p{3.0cm}|}
\hline
\textbf{Code} & \textbf{Thresholds} \\ \hline
Shor (3 rep X $ \times$ 3 rep Z) $\times$ 5 rep Z & 0.06360871676 \\ \hline
Shor (3 rep X $ \times$ 3 rep Z) $\times$ 4 rep Z & 0.06358111943 \\ \hline
Shor (3 rep X $ \times$ 3 rep Z) $\times$ 3 rep Z & 0.06352919792 \\ \hline
\red{Hashing} & \red{0.0630965616} \\ \hline
5-qubit $\times$ 5 rep & 0.06302575933 \\ \hline
5-qubit $\times$ 4 rep & 0.06302519416 \\ \hline
5-qubit $\times$ 3 rep & 0.06302258398 \\ \hline
Biased 9-qubit $\times$ 3 rep Z & 0.062809473853 \\ \hline
Biased 9-qubit $\times$ 3 rep X & 0.062726557443 \\ \hline
[[4,2,2]] $\times$ 4 rep & 0.06270942 \\ \hline
[[4,2,2]] $\times$ 3 rep & 0.06264641 \\ \hline
7-qubit $\times$ 3 rep & 0.06260170793 \\ \hline
7-qubit $\times$ 4 rep & 0.06257099182 \\ \hline
7-qubit $\times$ 5 rep & 0.062537073 \\ \hline
\end{tabular}
\caption{Error Thresholds for stabilizer $\times$ repetition codes (depolarizing channel). Note that in these examples when the first code is not the repetition code, then concatenation does not help do better than hashing.}
\label{table5}
\end{table}

\begin{figure}
    \centering
    \includegraphics[width=0.95\linewidth]{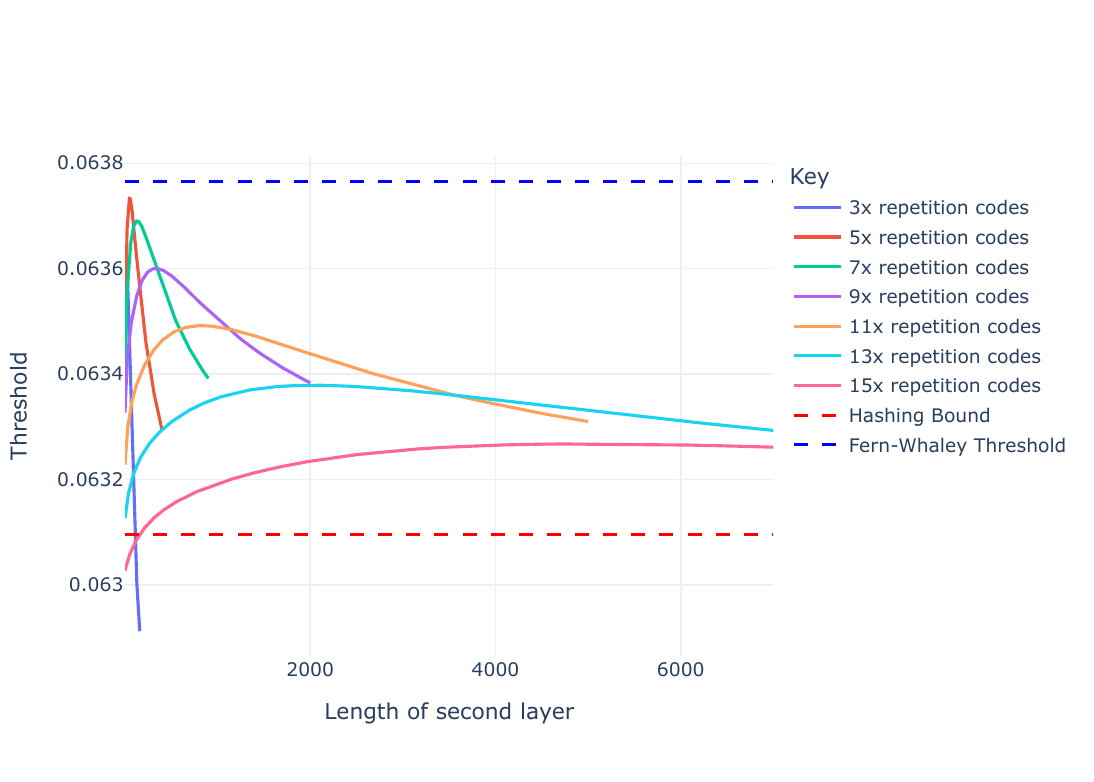}
    \caption{Performance estimates of long concatenated repetition codes for depolarizing channel. Note that the horizontal axis is the length of second layer of repetition code.}
    \label{depolarizingplot}
\end{figure}

We conclude the discussion in this section by studying the performance of concatenated repetition codes. For small sized repetition codes, we compute the threshold exactly, some of these data points were also reported in \cite{fern}. The exact computations are summarized in the Table \ref{table6}. Furthermore, using the results in Section \ref{weightenumerators} we were able to compute an estimate of the threshold for concatenations of long repetition codes (up to 15 $\times$ 7000). To the best of our knowledge, these search spaces have not been explored before. The thresholds are plotted in Figure \ref{depolarizingplot}. We see that on increasing the length of the first repetition codes, the threshold consistently gets worse (except when going from 3 rep to 5 rep), and the highest threshold in all the lengths we tried was given by the 5 $\times$ 51 repetition code. On fixing the length of the first repetition code and varying the length of the second repetition code, the threshold improves up to a certain length and then starts to get worse again. These results give several new examples of concatenated repetition codes which perform better than hashing. They further strengthen the evidence towards the belief that using longer repetition codes should not lead to better thresholds.

All the threshold improvements in previous works have the first code as a 5-repetition code. One reason to do this is that for a single layer the 5-repetition code gives the best threshold. This motivates a natural question: is it possible to find a pair of codes $A, B$ to concatenate, such that the threshold for Code $A~\times$ Code $B$ is better than that of 5-rep $\times$ Code $B$, even though Code $A$ by itself has a worse threshold than 5-rep. We see that it is indeed possible to find such codes. For example, with a single layer, 7-repetition has a better threshold than 3-repetition, but 3 $\times$ 7-rep has a better threshold than 7 $\times$ 7-rep, see Table \ref{table6}. For longer lengths, 5-rep has a better threshold than 7-rep, but 7 $\times$ 95-rep has a better threshold than 5 $\times$ 95-rep, see Figure \ref{depolarizingplot}.
\begin{table}
\centering
\small
\begin{tabular}{|l|>{\ttfamily}l|>{\ttfamily}l|>{\ttfamily}l|>{\ttfamily}l|}
\hline
\textbf{Outer $\backslash$ Inner} 
& \textbf{3 rep} 
& \textbf{4 rep} 
& \textbf{5 rep} 
& \textbf{7 rep} \\
\hline
\textbf{3 rep} 
& 0.06335987939
& 0.06338621397 
& 0.06341736592 
& 0.06347859503 \\
\hline
\textbf{4 rep} 
& 0.06328107294 
& 0.06327018271 
& 0.06325722134 
& 0.063226464487 \\
\hline
\textbf{5 rep} 
& 0.06349265709 
& 0.06350743142 
& 0.06352047426 
& 0.06354354564 \\
\hline
\textbf{7 rep} 
& 0.06344003002 
& 0.06345122519 
& 0.06346101398 
& 0.063477579990 \\
\hline
\textbf{No encoding} 
& 0.06337664297 
& 0.06329834885 
& 0.06345202939 
& 0.06341083749 \\
\hline
\end{tabular}
\caption{Concatenated repetition code thresholds (exact, depolarizing channel).}
\label{table6}
\end{table}

\begin{figure}
    \centering
    \includegraphics[width=0.95\linewidth]{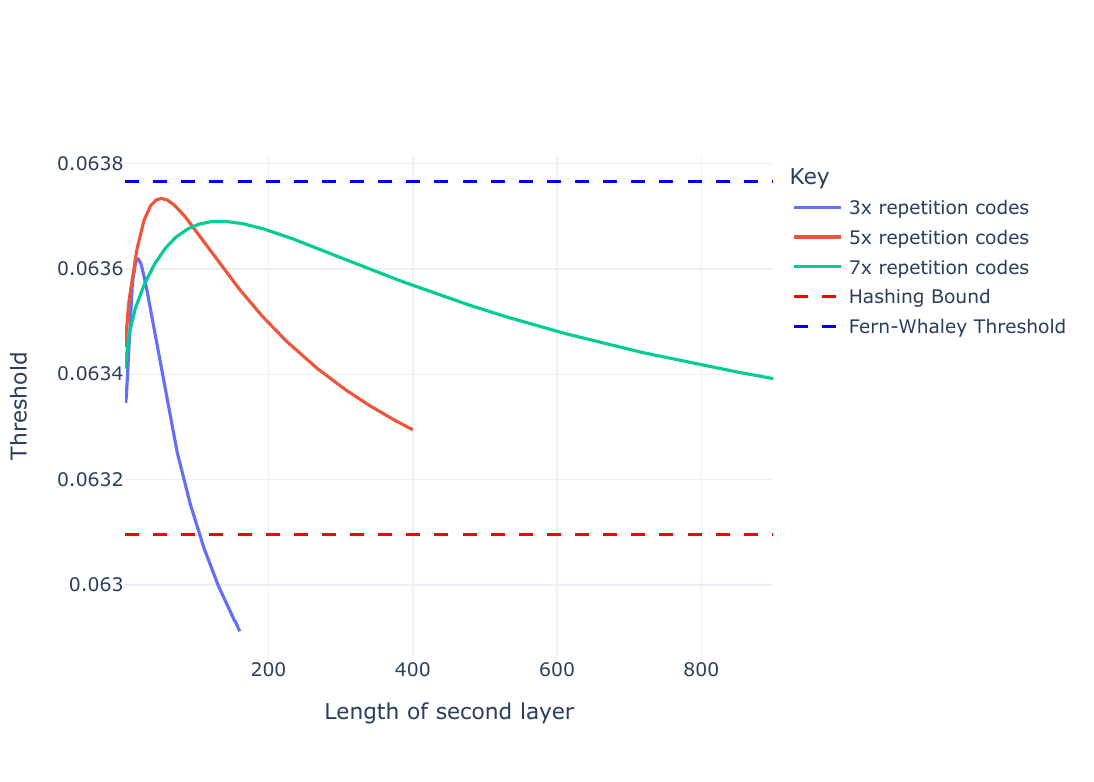}
    \caption{Zooming into threshold estimates for 3x, 5x, 7x concatenated repetition codes for depolarizing channel. Note that the horizontal axis is the length of second layer of repetition code.}
    \label{depolarizingzoomplot}
\end{figure}

\subsection{The Independent X-Z Channel}
In this section, we study the threshold problem for the Independent X-Z channel with parameters $((1-p)^2, p(1-p), p^2, p(1-p))$. This channel is also called BB-84 channel. One general observation to make right at the start is that the behavior of the code threshold for this channel is similar to that of depolarizing channel. This is somewhat surprising because given that the probability of the Y error in this channel is very small so one would expect it to be qualitatively similar to the 2 Pauli channel. However, as we will see later in Section \ref{twopauli}, this is not true. To start studying the threshold numerically we begin by computing the threshold for this channel when a single layer stabilizer code is used to encode the state. The thresholds are reported in Table \ref{table7}. Once again, the only codes which perform better than hashing are the repetition codes.

\begin{table}
\centering
\small
\begin{minipage}[t]{0.48\textwidth}
\centering
\begin{tabular}{|l|>{\ttfamily}p{3.0cm}|}
\hline
\textbf{Code} & \textbf{Threshold} \\
\hline
7 rep & 0.1121074112 \\
\hline
5 rep & 0.1121042613 \\
\hline
3 rep & 0.1116520369 \\
\hline
Shor code (9-qubit) & 0.1116273012 \\
\hline
4 rep & 0.1116162734 \\
\hline
\red{Hashing} & \red{0.1100278644} \\
\hline
Biased 13-qubit & 0.1095450642 \\
\hline
$[[11, 1, 5]]$ & 0.1095144549 \\
\hline
Biased 9-qubit & 0.1094880471 \\
\hline
\end{tabular}
\end{minipage}
\hfill
\begin{minipage}[t]{0.48\textwidth}
\centering
\begin{tabular}{|l|>{\ttfamily}p{3.0cm}|}
\hline
\textbf{Code} & \textbf{Threshold} \\
\hline
5 qubit & 0.1094667453 \\
\hline
7 qubit & 0.1094285595 \\
\hline
CD Steane & 0.109428519548 \\
\hline
[[6,1,3]] H & 0.109375692549 \\
\hline
Tailored $[[7, 1, 3]]$ H & 0.1093749886 \\
\hline
$[[4,2,2]]$ & 0.1092750072 \\
\hline
13-qubit cyclic code & 0.109176800515 \\
\hline
$[[8,2,2]]$ & 0.1086302482 \\
\hline
SCF H & 0.107975087086 \\
\hline
\end{tabular}
\end{minipage}
\vspace{0.3cm}
\caption{Error thresholds for various stabilizer codes (independent X-Z channel). The only non-concatenated codes which do better than random hashing are repetition codes.}
\label{table7}
\end{table}

\begin{table}
\centering
\footnotesize
\begin{minipage}[t]{0.49\textwidth}
\centering
\begin{tabular}{|l|>{\ttfamily}p{3.0cm}|}
\hline
\textbf{Code} & \textbf{Threshold} \\ \hline
7 rep X $\times$ Tailored [[7,1,3]] H & 0.1121999678 \\ \hline
5 rep Z $\times$ Biased 9-qubit & 0.112199245744 \\ \hline
5 rep X $\times$ Tailored [[7,1,3]] H & 0.112195710485 \\ \hline
5 rep Z $\times$ Tailored [[7,1,3]] H & 0.11219113341 \\ \hline
7 rep $\times$ 5 qubit & 0.1121760184 \\ \hline
5 rep $\times$ 5 qubit & 0.1121740643 \\ \hline
7 rep X $\times$ [[6,1,3]] H & 0.112160908299 \\ \hline
5 rep X $\times$ [[6,1,3]] H & 0.112147430289 \\ \hline
7 rep $\times$ CD Steane H & 0.1121454277 \\ \hline
5 rep $\times$ CD Steane H & 0.1121390516 \\ \hline
\red{7 rep} & \red{0.1121074112} \\ \hline
\red{5 rep} & \red{0.1121042613} \\ \hline
4 rep $\times$ SCF H & 0.1118931344 \\ \hline
3 rep $\times$ 7 qubit & 0.1118663697 \\ \hline
4 rep $\times$ $[[4,2,2]]$ & 0.1118332576 \\ \hline
4 rep Z $\times$ [[6,1,3]] H & 0.111824435304 \\ \hline
5 rep Z $\times$ [[6,1,3]] H & 0.111816604543 \\ \hline
5 rep $\times$ $[[4,2,2]]$ & 0.1118009187 \\ \hline
3 rep $\times$ SCF H & 0.1117956372 \\ \hline
\end{tabular}
\end{minipage}
\hfill
\begin{minipage}[t]{0.49\textwidth}
\centering
\begin{tabular}{|l|>{\ttfamily}p{3.0cm}|}
\hline
\textbf{Code} & \textbf{Threshold} \\ \hline
5 rep $\times$ SCF H & 0.1117609539 \\ \hline
4 rep $\times$ 7 qubit & 0.1116669006 \\ \hline
3 rep & 0.1116520369 \\ \hline
4 rep X $\times$ [[6,1,3]] H & 0.111625259524 \\ \hline
3 rep $\times$ $[[4,2,2]]$ & 0.1116243095 \\ \hline
4 rep $\times$ CD Steane H & 0.1115988745 \\ \hline
4 rep $\times$ 5 qubit & 0.1115748657 \\ \hline
4 rep X $\times$ Tailored [[7,1,3]] H & 0.111563933 \\ \hline
3 rep Z $\times$ [[6,1,3]] H & 0.11156195318 \\ \hline
3 rep $\times$ 5 qubit & 0.111530293 \\ \hline
3 rep Z $\times$ Tailored [[7,1,3]] H & 0.11152247055 \\ \hline
3 rep X $\times$ Tailored [[7,1,3]] H & 0.111519718682 \\ \hline
3 rep Z $\times$ Biased 9-qubit & 0.111516956 \\ \hline
3 rep $\times$ CD Steane H & 0.1114983542 \\ \hline
7 rep Z $\times$ [[6,1,3]] H & 0.111479474652 \\ \hline
3 rep X $\times$ [[6,1,3]] H & 0.111457260799 \\ \hline
5 rep $\times$ 7 qubit & 0.1114316579 \\ \hline
7 rep $\times$ SCF H & 0.1113603815 \\ \hline
\red{Hashing} & \red{0.1100278644} \\ \hline
\end{tabular}
\end{minipage}

\caption{Repetition codes concatenated with stabilizer codes (independent X-Z channel). Improved thresholds when concatenating with holographic codes and biased 9-qubit code.}
\label{table8}
\end{table}

\begin{figure}
    \centering
    \includegraphics[width=0.80\linewidth]{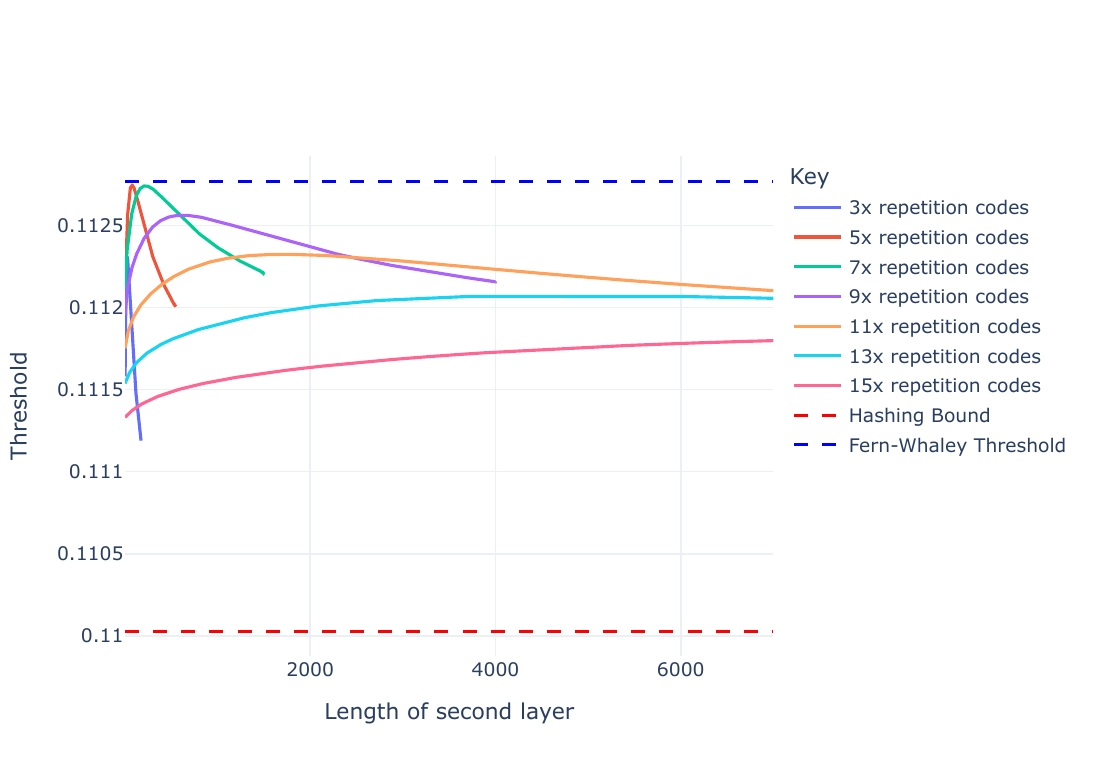}
    \caption{Thresholds estimates for long concatenated repetition codes for independent X-Z channel. Note that the horizontal axis is the length of second layer of repetition code.}
    \label{fig:indxzplot}
\end{figure}
\begin{figure}
    \centering
    \includegraphics[width=0.80\linewidth]{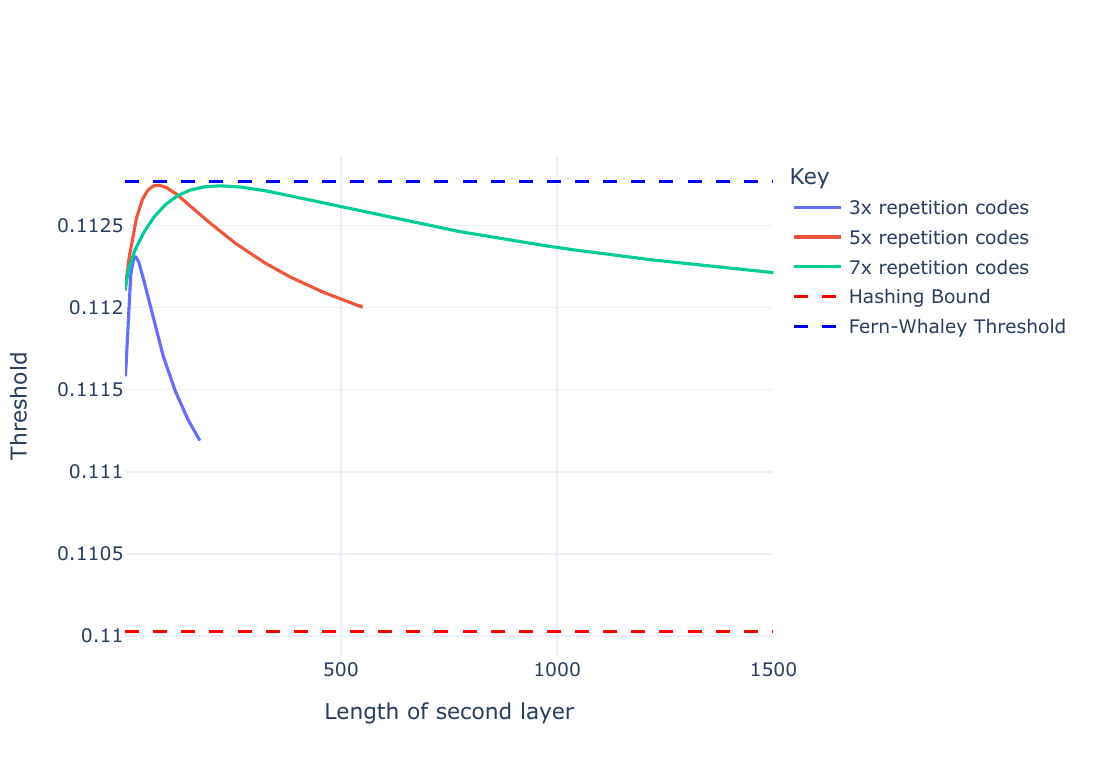}
    \caption{Zooming into threshold estimates for 3x, 5x, 7x concatenated repetition codes for independent X-Z channel. Note that the horizontal axis is the length of second layer of repetition code.}
\end{figure}
We next studied the thresholds for repetition codes concatenated with small stabilizer codes (including the holographic codes and biased codes discussed in the previous section). The thresholds are reported in Table \ref{table8}. Note that again concatenating repetition codes with holographic codes and the biased 9-qubit code improves the threshold. The best performance for a single layer is for 7-repetition and 5-repetition codes, and the best holographic codes to concatenate are the tailored [[7,1,3]], 5-qubit, [[6,1,3]] and CD Steane codes. Concatenating 5-rep with biased 9-qubit code also shows a significant improvement in threshold. Note that these values have not been previously reported.

\begin{table}
\centering
\footnotesize
\begin{tabular}{|l|>{\ttfamily}l|>{\ttfamily}l|>{\ttfamily}l|>{\ttfamily}l|}
\hline
\textbf{Outer $\backslash$ Inner} & \textbf{3 rep} & \textbf{4 rep} & \textbf{5 rep} & \textbf{7 rep} \\
\hline
\textbf{3 rep} & 0.1116273012 & 0.1116758826 & 0.1117431391 & 0.111872313296 \\
\hline
\textbf{4 rep} & 0.111549439  & 0.1115223945 & 0.1114951008 & 0.1114356246 \\
\hline
\textbf{5 rep} & 0.1121585875 & 0.1121812393 & 0.1122021756 & 0.1121911334 \\
\hline
\textbf{7 rep} & 0.1121487919 & 0.1121656721 & 0.1121808508 & 0.112207468445 \\
\hline
\textbf{No encoding} & 0.1116520369 & 0.1116162734 & 0.1121042613 & 0.1121074112 \\
\hline
\end{tabular}
\caption{Concatenated repetition code thresholds (exact, independent X-Z channel).}
\label{table9}
\end{table}

We estimate the thresholds for concatenated repetition codes of long lengths for the independent X-Z channel. All of them do better than hashing, and many of them do better than 7-repetition code. While none of them do better than the threshold in \cite{fern}, the peaks of 5-repetition and 7-repetition concatenations come very close to it. These peak values $p = 0.1127458434,~p = 0.1127420360$ respectively, were also reported in \cite{fern}, but we would like to note that our work is the first one to compute the thresholds for very long repetition codes for the independent X-Z channel and hence give many new examples of non-additivity. The general trend for these repetition codes is similar to that of the depolarizing channel, that is, the threshold becomes worse as the length increases.
\subsection{The 2-Pauli Channel}
\label{twopauli}
\begin{figure}
    \centering
    \includegraphics[width=0.85\linewidth]{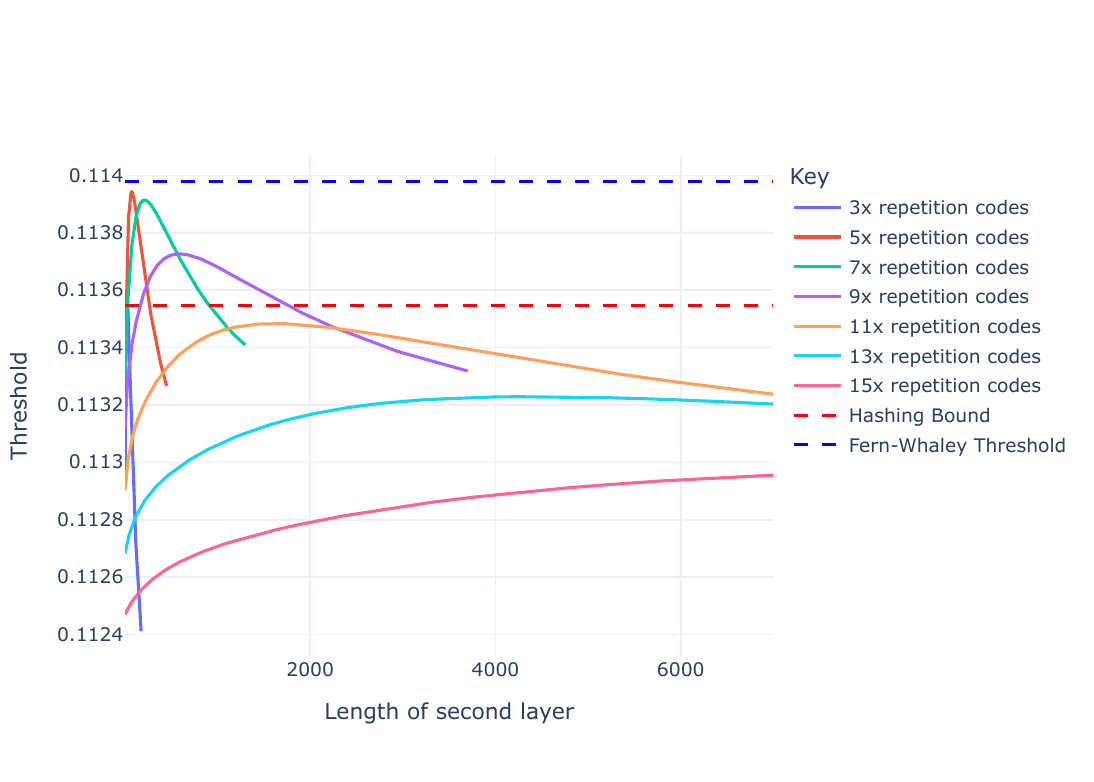}
    \caption{Thresholds estimates for long concatenated repetition codes for 2-Pauli channel. Note that the horizontal axis is the length of second layer of repetition code.}
    \label{fig:twopauliplot}
\end{figure}
\begin{figure}
    \centering
    \includegraphics[width=0.80\linewidth]{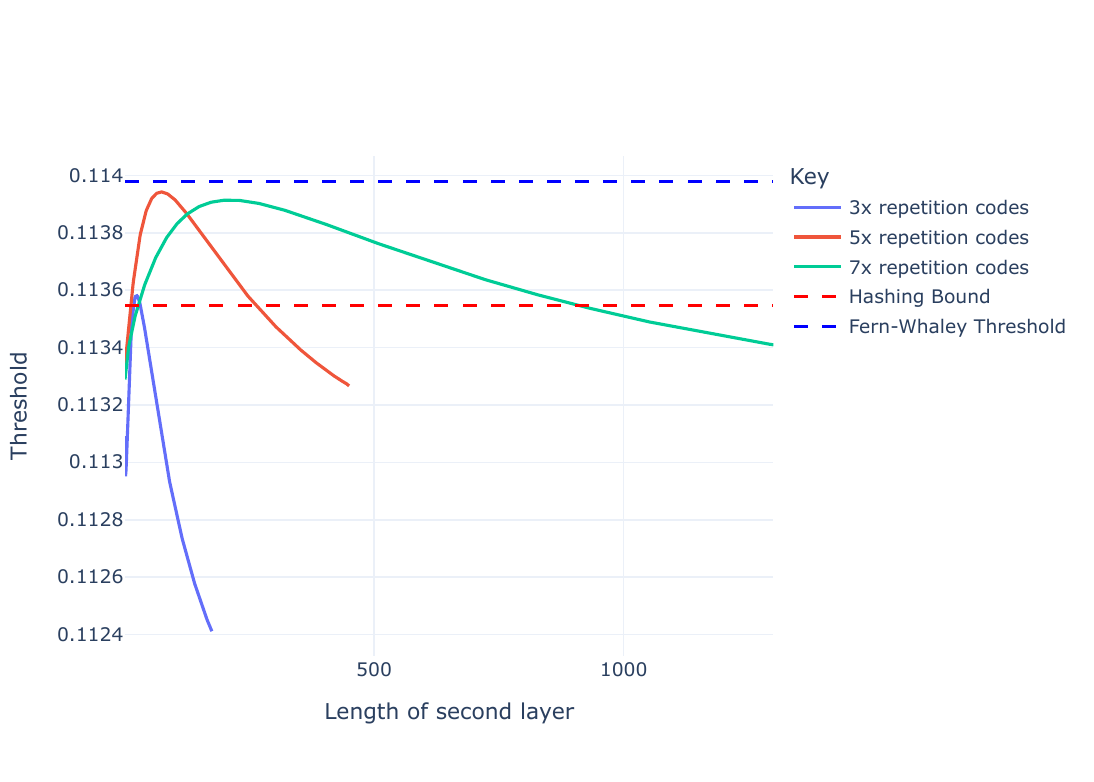}
    \caption{Zooming into threshold estimates for 3x, 5x, 7x concatenated repetition codes for 2-Pauli channel. Note that the horizontal axis is the length of second layer of repetition code.}
\end{figure}
We estimate the thresholds for several concatenated repetition codes of long lengths and find that many of them outperform hashing. However, none of them do better than the best threshold in \cite{fern}. Note that the peak value obtained here for 5 repetition $\times$ 74 repetition was also reported (exactly) in \cite{fern}, but our work is the first to estimate the thresholds for the other long repetition codes. Furthermore, from the results in the previous sections, one might form the impression that non-additivity is primarily a small block-length phenomenon. However, we see that this is not true for the 2-Pauli channel.

In particular, we do not find any other small stabilizer code (or concatenations) that does better than hashing for the 2-Pauli channel. We report these results only to record the fact that these codes are useless from the perspective of non-additivity of this channel. Even though all the codes we tried are worse than hashing, the only times we see an improvement by concatenating codes is when we concatenate with either another repetition code or holographic codes. For a detailed table with the threshold values for small stabilizer codes, see the Github repository \cite{github}.\footnote{See file Code Thresholds.pdf}

\section{Channel Optimizations}
In this section, we flip the problem: instead of trying to find codes which give better threshold for a specific channel, we fix the code and vary the channel parameters with the objective of maximizing the coherent information at the hashing point of the corresponding channel. The results of this optimization for single-layer codes is summarized in Table \ref{table10} and for concatenated codes in Table \ref{table11}. In particular, let $(c_X, c_Y, c_Z)$ refer to the $X, Y, Z$ coefficients respectively in the Tables \ref{table10}, \ref{table11}. Then the optimized channel has parameters $(1-p, c_Xp, c_Yp, c_Zp)$, with total error $p$.\footnote{We optimize for coefficients in the range (0.0001, 1).} The entry in the hashing point column tells us the value of $p_{hash}$ at which a random stabilizer code gives capacity $0$ for this channel, i.e, when the entropy of the errors of the channel is $1$. The entry in the non-additivity column tells us the capacity achieved at $p_{hash}$ when using a specific code before random hashing. Note that we use these values to highlight general trends.
\begin{table}
\centering
\small
\renewcommand{\arraystretch}{0.85}
\setlength{\extrarowheight}{0.3ex}
\setlength{\tabcolsep}{4pt}
\begin{tabular}{|l|>{\ttfamily}l|>{\ttfamily}l|>{\ttfamily}l|>{\ttfamily}l|>{\ttfamily}l|}
\hline
\textbf{Code}
& \textbf{X coefficient}
& \textbf{Y coefficient}
& \textbf{Z coefficient}
& \textbf{Non-additivity}
& \textbf{Hashing point} \\
\hline
4 rep & 0.06609142 & 0.91039291 & 0.02351567 & 0.012959633 & 0.2810011867 \\
\hline
3 rep & 0.08519012 & 0.0277214 & 0.88708848 & 0.01274328527 & 0.2683777284 \\
\hline
5 rep & 0.05116499 & 0.0208338 & 0.92800121 & 0.01259428267 & 0.2928465458 \\
\hline
7 rep & 0.0374894 & 0.01589898 & 0.94661162 & 0.0114603097 & 0.3093972342 \\
\hline
5 qubit & 0.02058418 & 0.0205851 & 0.95883071 & 0.008869175026 & 0.3223408279 \\
\hline
Tailored $$[[7, 1, 3]]$$ H & 0.01496641 & 0.01825069 & 0.96678289 & 0.008533622316 & 0.3338306974 \\
\hline
Shor code (9-qubit) & 0.01488571 & 0.96882848 & 0.01628581 & 0.008297933843 & 0.3370935218 \\
\hline
Biased 9-qubit & 0.01216847 & 0.01216844 & 0.9756631 & 0.007489070428 & 0.349643758 \\
\hline
13-qubit cyclic code & 0.98030287 & 0.00961704 & 0.01008009 & 0.00672953481 & 0.3599226452 \\
\hline
Biased 13-qubit & 0.00862734 & 0.00862735 & 0.98274531 & 0.00637290432 & 0.3661221308 \\
\hline
$$[[6, 1, 3]]$$ H & 0.95454889 & 0.01609587 & 0.02935524 & 0.005887070178 & 0.3175941557 \\
\hline
CD Steane H & 0.02960147 & 0.01339475 & 0.95700378 & 0.002980713554 & 0.321036373 \\
\hline
$$[[11, 1, 5]]$$ & 0.01404342 & 0.01172789 & 0.97422869 & 0.00123245788 & 0.3468338078 \\
\hline
7 qubit & 0.47099554 & 0.4707756 & 0.05822885 & \red{-0.00086150789} & 0.2079492418 \\
\hline
SCF H & 0.41043731 & 0.52427592 & 0.06528677 & \red{-0.0021075336} & 0.2072578017 \\
\hline
Toric code $$[[8, 2, 2]]$$ & 0.07057851 & 0.49868546 & 0.43073603 & \red{-0.00233821885} & 0.205903352 \\
\hline
[[4,2,2]] & 0.47089171 & 0.05821602 & 0.47089227 & \red{-0.00307897972} & 0.2079497163 \\
\hline
\end{tabular}
\caption{The optimized channel has parameters $(1-p, c_Xp, c_Yp, c_Zp)$, with total error $p$. The entry in the hashing point column tells us the value of $p_{hash}$ at which a random stabilizer code gives capacity $0$ for this channel. The entry in the non-additivity column tells us the capacity achieved at $p_{hash}$.}
\label{table10}
\end{table}

We arrange the codes in the tables so that we first put codes which show maximal coherent information for channels which are very highly biased towards a single type of error. The codes in the bottom of the table show maximal coherent information for channels which have roughly equal probability of two types of errors, and some small probability of the third type of error. Note that these are the codes which do not show non-additivity even after channel optimization (7-qubit Steane, [[4,2,2]], Toric code [[8,2,2]] and SCF). These codes are also at the bottom of the threshold table for both depolarizing channel and independent X-Z channel, and they do significantly worse than hashing. We also see the coherent information for longer codes in Table \ref{table10} is significantly worse than that for the shorter repetition codes. As seen in \ref{table11}, concatenating other codes after the first layer of repetition codes makes the coherent information worse, possibly due to the longer length. Results in these tables might inform which codes to pick in a biased noise regime, and may be relevant for both error correction and capacity.
\begin{table}
\centering
\small
\renewcommand{\arraystretch}{0.85}
\setlength{\extrarowheight}{0.3ex}
\setlength{\tabcolsep}{4pt}
\begin{tabular}{|l|>{\ttfamily}l|>{\ttfamily}l|>{\ttfamily}l|>{\ttfamily}l|>{\ttfamily}l|}
\hline
\textbf{Concatenated Code} 
& \textbf{X coefficient} 
& \textbf{Y coefficient} 
& \textbf{Z coefficient} 
& \textbf{Non-additivity} 
& \textbf{Hashing point} \\
\hline
3 rep X $\times$ 5 rep Z 
& 0.00840665 
& 0.980756 
& 0.01083735 
& 0.00688476977 
& 0.3611050233 \\
\hline
5 rep X $\times$ 3 rep Z 
& 0.01454433 
& 0.9761731	 
& 0.00928257 
& 0.00673676503	
& 0.3509685235 \\
\hline
3 rep X $\times$ 5 qubit 
& 0.01256495 
& 0.00740623 
& 0.98002881 
& 0.00638679074 
& 0.3596135389 \\
\hline
3 rep Z $\times$ 7 rep X 
& 0.00827083 
& 0.98581108 
& 0.00591809 
& 0.00586846919 
& 0.3750748571 \\
\hline
3 rep Z $\times$ Shor 9 qubit
& 0.01215838
& 0.94775012
& 0.04009151
& 0.0051414728
& 0.3116335074 \\
\hline
5 rep X $\times$ 5 rep Z	
& 0.00584507	
& 0.98816151
& 0.00599342	
& 0.0050847667	
& 0.38276095094\\
\hline
5 rep Z $\times$ 5 qubit	
& 0.00519623	
& 0.9855565	
& 0.00924727	
& 0.0049542287	
& 0.3745254396 \\
\hline
3 rep X $\times$ Biased 9-qubit 
& 0.00550621 
& 0.0044036 
& 0.99009019 
& 0.00440877222 
& 0.39012831 \\
\hline
3 rep X $\times$ 7 qubit 
& 0.13300218 
& 0.83074256 
& 0.03625527 
& 0.002980809573 
& 0.2470615767 \\
\hline
3 rep Z $\times$ Biased 9-qubit 
& 0.02340259 
& 0.89070293 
& 0.08589448 
& 0.002962751111 
& 0.2710045173 \\
\hline
5 rep Z $\times$ 7 rep X	
& 0.02967107	
& 0.77531338	
& 0.19501555	
& 0.002175342298	
& 0.2361625619 \\
\hline
\end{tabular}
\caption{Channel optimization for selected concatenated codes. Here the columns have the same meaning as in the previous table.}
\label{table11}
\end{table}

\section{Discussion and Outlook}
The goal of this paper was to understand what properties of stabilizer codes lend themselves to better thresholds or non-additivity. To this end, we tried several large repetition codes and new stabilizer code concatenations. The choice of the codes that we picked to run was based on a few factors:
\begin{itemize}
\item Degeneracy: We analyzed and ran large repetition codes in order to understand the extent to which their degeneracy improves the threshold. We found that as the lengths of the concatenated repetition codes increases, the threshold becomes worse. This gives rise to the following question: We know that degeneracy is necessary to show non-additivity \cite{shor1996quantumerrorcorrectingcodesneed} but how to quantify the \emph{amount} of degeneracy as a function of the length of the code is unclear. What complicates the situation further is that in the context of capacity we care about degeneracy of a particular code with respect to typical errors and this is conceptually different from how we define degeneracy for error correcting codes which is usually done in the context of adversarial low weight errors.
\item Performance as error correcting codes: We picked codes such as the 5-qubit code, 7-qubit code, Toric code, [[4,2,2]] code and 11-qubit code for their interesting error correction properties. One natural metric for this is the distance of the code, since higher distance should allow the code to correct more errors. However as we can see from the thresholds, these codes do not perform better than just random stabilizer codes, whereas the repetition codes with distance 1 outperform random stabilizer codes. Further, concatenating these codes with repetition codes in the second layer performs worse than random stabilizer codes. To see how distance is conceptually different from capacity, we note that the distance bounds of \cite{rains2, selfdual} show that no stabilizer code can correct errors of weight more than $\lfloor\frac{n+1}{6}\rfloor$, which corresponds to an error rate of roughly $16.6\%$. However, we see that using a random code we can achieve capacity up to error rates of roughly $18.9\%$. This difference occurs because the codes are designed to correct even errors of weight $1$, but these errors are highly unlikely to occur in the capacity setup, so we can ignore them. Given this knowledge about degeneracy and error correction, a good test could be to study the performance of LDPC codes, which have low-weight stabilizers (and hence good for degeneracy), as well as high distance.
\item Codes for Biased Channels: Many of the codes we considered were tailored for showing better thresholds for biased channels. Since the effective channels after the repetition codes are biased, we concatenate repetition codes with these code families and study the threshold. In this case, our intuition turns out to be correct and we see improved thresholds. Note also that all the biased codes we considered have non-trivial distance. \cite{fern} concatenated multiple layers of the 5-qubit code to get the best threshold result for depolarizing channel, which motivates the question of whether the code distance starts making a difference once the channel is suitably biased. Moreover, we do not try concatenating a very large number of layers of codes, and it remains open to see whether concatenating multiple layers of any of the biased codes can give better thresholds.
\end{itemize}
We now make some remarks about other small experiments we carried out during our exploration of this problem. Since in the first layer distance of the code does not seem to matter, we tried generating random stabilizer codes up to $n=7$ and calculate the thresholds for the resulting code samples. We found several examples of non-additivity both for depolarizing and independent X-Z channel, but none for 2-Pauli. In order to see where we stop seeing non-additivity while approaching the 2-Pauli channel, we tried to study this channel in the limit. We found that numerically, on introducing a very small amount of Y error probability, small repetition codes start showing non-additivity. However, these limiting channels are still very far from independent X-Z type channel.

We have very few examples of stabilizer codes (needs very long length) which show non-additivity for the 2-Pauli channel. Recently, \cite{bhalerao2025improvingquantumcommunicationrates} showed that using permutation invariant codes of small lengths shows non-additivity for both 2-Pauli and independent X-Z channel (but not for depolarizing channel). An immediate conclusion one could draw is that permutation-invariant codes behave similarly for the two channels, which is in contrast with the case for stabilizer codes. In the latter case, we instead see similar behavior for independent X-Z and depolarizing channel thresholds. This motivates the question of trying to better understand the connection between the stabilizer code approach and the permutation-invariant code approach. A first step towards building a bridge between these would be to translate the coset weight enumerator framework in the language of representation theory.
\section*{Acknowledgments}
AA is supported in part by a Cheriton Graduate Scholarship from the School of Computer Science at the University of Waterloo. ARK acknowledges the support of the CryptoWorks 21 program. AB was supported by the Mitacs Globalink Research Internship and the Undergraduate School on Experimental Quantum Information Processing. Research at Perimeter Institute and IQC is supported by the Government of Canada through Innovation, Science and Economic Development Canada, and by the Province of Ontario through the Ministry of Research, Innovation and Science. 

This research was enabled in part by support provided by Compute Ontario (\url{www.computeontario.ca}) and the Digital Research Alliance of Canada (\url{www.alliancecan.ca}). The AWS cloud was also used for computing facilities.

We also acknowledge the support of the Natural Sciences and Engineering Research Council of Canada (NSERC). 
AA and ARK received partial support from the NSERC Alliance Consortia Quantum grants (ALLRP 578455-22). DL is supported under RGPIN-2024-03823, LS is supported under RGPIN-2025-04875, and GS is supported under NSERC-NSF alliance grant ALLRP-586858-2023 and NSERC Discovery grant RGPIN-2025-02094.

\printbibliography
\appendix
\clearpage
\section{Stabilizer Generators and Logical Operators}
\label{app}
\begin{table}[H]
\centering
\small
\renewcommand{\arraystretch}{1.3} 
\begin{tabularx}{\textwidth}{|l|X|p{4.5cm}|}
\hline
\textbf{Code} & \textbf{Stabilizer Generators} & \textbf{Logical Operators} \\ \hline
3 Rep (X) & \texttt{XXI}, \texttt{XIX} & \texttt{XII}, \texttt{ZZZ} \\ \hline
3 Rep (Z) & \texttt{ZZI}, \texttt{ZIZ} & \texttt{XXX}, \texttt{ZII} \\ \hline
4 Rep (Z) & \texttt{ZZII}, \texttt{ZIZI}, \texttt{ZIIZ} & \texttt{XXXX}, \texttt{ZIII} \\ \hline
5 Rep (Z) & \texttt{ZZIII}, \texttt{ZIZII}, \texttt{ZIIZI}, \texttt{ZIIIZ} & \texttt{XXXXX}, \texttt{ZIIII} \\ \hline
7 Rep (X) & \texttt{XXIIIII}, \texttt{XIXIIII}, \texttt{XIIXIII}, \texttt{XIIIXII}, \texttt{XIIIIXI}, \texttt{XIIIIIX} & \texttt{XIIIIII}, \texttt{ZZZZZZZ} \\ \hline
5-Qubit & \texttt{XZZXI}, \texttt{IXZZX}, \texttt{XIXZZ}, \texttt{ZXIXZ} & \texttt{XXXXX}, \texttt{ZZZZZ} \\ \hline
7-Qubit (Steane) & \texttt{IIIXXXX}, \texttt{IXXIIXX}, \texttt{XIXIXIX}, \texttt{IIIZZZZ}, \texttt{IZZIIZZ}, \texttt{ZIZIZIZ} & \texttt{XXXXXXX}, \texttt{ZZZZZZZ} \\ \hline
Tailored [[7,1,3]] H & \texttt{XZIZXII}, \texttt{IXZIZXI}, \texttt{IIXZIZX}, \texttt{XIIXZIZ}, \texttt{ZXIIXZI}, \texttt{IZXIIXZ}, \texttt{ZIZXIIX} & \texttt{XXXXXXX}, \texttt{ZZZZZZZ} \\ \hline
[[6,1,3]] H & \texttt{ZIZIII}, \texttt{XZYYXI}, \texttt{XXXXZI}, \texttt{IZZXIX}, \texttt{XYXYIZ} & \texttt{XZXZII}, \texttt{XYYXII} \\ \hline
CD Steane H & \texttt{XZZIIIX}, \texttt{XIZXZII}, \texttt{XIIIZZX}, \texttt{ZXXIIIZ}, \texttt{ZIXZXII}, \texttt{ZIIIXXZ} & \texttt{XZZXZZX}, \texttt{ZXXZXXZ} \\ \hline
SCF H & \texttt{XXIXI}, \texttt{IIXXX}, \texttt{ZIZZI}, \texttt{IZIZZ} & \texttt{XIXII}, \texttt{IIZIZ} \\ \hline
Shor & \texttt{ZZIIIIIII}, \texttt{ZIZIIIIII}, \texttt{IIIZZIIII}, \texttt{IIIZIZIII}, \texttt{IIIIIIZZI}, \texttt{IIIIIIZIZ}, \texttt{XXXXXXIII}, \texttt{XXXIIIXXX} & \texttt{XXXXXXXXX}, \texttt{ZZZZZZZZZ} \\ \hline
11-Qubit & \texttt{ZZZZZZIIIII}, \texttt{XXXXXXIIIII}, \texttt{IIIZXYYYYXZ}, \texttt{IIIXYZZZZYX}, \texttt{ZYXIIIZYXII}, \texttt{XZYIIIXZYII}, \texttt{IIIZYXXYZII}, \texttt{IIIXZYZXYII}, \texttt{ZXYIIIZZZXY}, \texttt{YZXIIIYYYZX} & \texttt{IIIIIIXXXXX}, \texttt{IIIIIIZZZZZ} \\ \hline
13-Qubit & \texttt{XIZZIXIIIIIII}, \texttt{IXIZZIXIIIIII}, \texttt{IIXIZZIXIIIII}, \texttt{IIIXIZZIXIIII}, \texttt{IIIIXIZZIXIII}, \texttt{IIIIIXIZZIXII}, \texttt{IIIIIIXIZZIXI}, \texttt{IIIIIIIXIZZIX}, \texttt{XIIIIIIIXIZZI}, \texttt{IXIIIIIIIXIZZ}, \texttt{ZIXIIIIIIIXIZ}, \texttt{ZZIXIIIIIIIXI} & \texttt{XXXXXXXXXXXXX}, \texttt{ZZZZZZZZZZZZZ} \\ \hline
Biased 9-Qubit & \texttt{ZZIZIZIXY}, \texttt{XZZIZZIIX}, \texttt{IXZIIZZZY}, \texttt{IZXZZIZIY}, \texttt{ZIZXZIIZY}, \texttt{ZZIIXIZZX}, \texttt{ZIZZIXZIX}, \texttt{IIIZZZXZX} & \texttt{IIIIXXIIX}, \texttt{ZZZZZZZZZ} \\ \hline
Biased 13-Qubit & \texttt{XZIZZZIZIIIZX}, \texttt{IXZIZZZZZIIIY}, \texttt{ZZXZIIIZZZIIY}, \texttt{IIZXIZIZIZZZY}, \texttt{ZIZZXIZZIIZIX}, \texttt{IZIZZXZIIZZIY}, \texttt{IIZZZIXIZZIZX}, \texttt{ZIIIZZIXZZZIX}, \texttt{ZZZIZIIIXIZZY}, \texttt{IZIIIIZZZXZZX}, \texttt{ZIIZIZZIZIXZY}, \texttt{ZZZIIZZIIZIXX} & \texttt{XIIXXXIXIIXIX}, \texttt{ZZXXIXIXIIXII} \\ \hline
[[4,2,2]] & \texttt{XXXX}, \texttt{ZZZZ} & \texttt{XXII}, \texttt{IXXI}, \texttt{IZZI}, \texttt{ZZII} \\ \hline
Toric [[8,2,2]] & \texttt{XXIIXIXI}, \texttt{IIXXXIXI}, \texttt{XXIIIXIX}, \texttt{ZIZIZZII}, \texttt{IZIZZZII}, \texttt{ZIZIIIZZ} & \texttt{XIXIIIII}, \texttt{IIIIXXII}, \texttt{ZZIIIIII}, \texttt{IIIIZIZI} \\ \hline
\end{tabularx}
\label{quantumcodes}
\end{table}
\end{document}